# Explainable Artificial Intelligence Applications in Cyber Security: State-of-the-Art in Research

**ZHIBO ZHANG[1,2], HUSSAM AL HAMADI[1,2], (Senior Member, IEEE), ERNESTO DAMIANI[1,2], (Senior Member, IEEE), CHAN YEOB YEUN[1,2], (Senior Member, IEEE), and FATMA TAHER[3], (Senior Member, IEEE)**

[1]Center for Cyber-Physical Systems, Khalifa University, Abu Dhabi, United Arab Emirates
[2]Department of Electrical Engineering and Computer Science, Khalifa University, Abu Dhabi, United Arab Emirates
[3]College of Technological Innovation, Zayed University, Dubai, United Arab Emirates

Corresponding author: Zhibo Zhang (e-mail: qiuyuezhibo@gmail.com).

**ABSTRACT** This survey presents a comprehensive review of current literature on Explainable Artificial Intelligence (XAI) methods for cyber security applications. Due to the rapid development of Internet-connected systems and Artificial Intelligence in recent years, Artificial Intelligence including Machine Learning (ML) and Deep Learning (DL) has been widely utilized in the fields of cyber security including intrusion detection, malware detection, and spam filtering. However, although Artificial Intelligence-based approaches for the detection and defense of cyber attacks and threats are more advanced and efficient compared to the conventional signature-based and rule-based cyber security strategies, most ML-based techniques and DL-based techniques are deployed in the ‘‘black-box’’ manner, meaning that security experts and customers are unable to explain how such procedures reach particular conclusions. The deficiencies of transparencies and interpretability of existing Artificial Intelligence techniques would decrease human users’ confidence in the models utilized for the defense against cyber attacks, especially in current situations where cyber attacks become increasingly diverse and complicated. Therefore, it is essential to apply XAI in the establishment of cyber security models to create more explainable models while maintaining high accuracy and allowing human users to comprehend, trust, and manage the next generation of cyber defense mechanisms. Although there are papers reviewing Artificial Intelligence applications in cyber security areas and the vast literature on applying XAI in many fields including healthcare, financial services, and criminal justice, the surprising fact is that there are currently no survey research articles that concentrate on XAI applications in cyber security. Therefore, the motivation behind the survey is to bridge the research gap by presenting a detailed and up-to-date survey of XAI approaches applicable to issues in the cyber security field. Our work is the first to propose a clear roadmap for navigating the XAI literature in the context of applications in cyber security.

## I. INTRODUCTION

Cyber Security is the practice of securing networks, devices, and data against unauthorized access or illegal usage, as well as the art of maintaining information confidentiality, integrity, and availability [1], whereas cyber defensive mechanisms emerge at the application, network, host, and data levels [2]. As the Internet has become an essential tool in everyone's daily life, the number of systems linked to the Internet grows as well. The advancement of computer networks, servers, and mobile devices has significantly boosted Internet usage. However, the wide utilization of the Internet also tempts cyber attackers to develop more sophisticated and powerful cyber-attack methods for their benefit. It is noticeable that with the number of internet users worldwide increasing by 0.3 billion in 2021 compared with the previous year [3], global cyber attacks increased by 29% in 2021 according to the 2021 Cyber Trends Report [4]. In June of 2022, a cyberattack on a software business caused thousands of individuals in multiple states of the USA to lose their unemployment benefits and job-search help [5], which will lead to severe social instability during the COVID-19 pandemic. As a matter of fact, according to the report by the European Union Agency for Network and Information

Security (ENISA) [6], safe and trustworthy cyberspace is expected to become even more crucial in the new social and economic norms formed by the COVID-19 epidemic. These figures and events demonstrate the serious facts that the Internet and connected networks and devices have suffered more cybercriminals and cyber attacks nowadays.

Therefore, a stable and secure cyber security computer system must be established to ensure the information privacy, accessibility, and integrity transmitted within the Internet. Nevertheless, the conventional signature-based and rule-based cyber defensive mechanisms are facing challenges within the increasing quantities of information spread over the Internet [7]. On the other hand, cyber hackers are always striving to keep one step ahead of law enforcement by generating new, smart, and intricate attacking techniques and implementing technological advances including Artificial Intelligence to make their adversarial behaviors more sophisticated and efficient [8]. As a consequence, researchers in cyber security have begun to investigate Artificial Intelligence-based approaches especially ML and DL rather than traditional (non-AI) cybersecurity techniques including Game theory, Rate Control, and Autonomous systems to enhance the performance of cyber defensive systems.

Although Artificial Intelligence techniques, especially ML and DL algorithms could provide impressive performances on benchmark datasets in a number of cyber security domain applications such as Intrusion detection, spam e-mail filtering, Botnet detection, fraud detection, and malicious application identification [9], they can commit errors, some of which are more expensive than conventional cyber defensive approaches. On the other hand, cyber security developers have sometimes sought higher accuracy at the price of interpretability, making their models more intricate and difficult to grasp [10]. This lack of explainability has been disclosed by the European Union's General Data Protection Regulation, preserving the capacity to comprehend the logic behind an Artificial Intelligence algorithmic decision that negatively impacts individuals [11]. Accordingly, to be able to believe the decisions of cyber security systems, Artificial Intelligence must be transparent and interpretable. To satisfy these kinds of demands, several strategies have been proposed to make Artificial Intelligence decisions more intelligible to humans. And these explainable techniques are usually shortened as "XAI", which have already been implemented in many application domains such as healthcare, Natural Language Processing, and financial services [12]. And the objective of this research paper is to focus on the applications of XAI in different fields in the context of cyber security.

### A. RESEARCH MOTIVATION

Implementing Artificial Intelligence in applications of cyber security has been researched in recent years and many previous surveys reviewed the existing work in this field. On the other hand, the trends of applying XAI to provide more explainable and transparent services for areas including healthcare and image analysis are popular in research as well. However, to the best of our knowledge, although there are some other excellent survey papers available on the topics of XAI and cyber security independently, there is a lack of a comprehensive survey paper focusing on the review of solutions based on XAI across a wide variety of cyber security applications. This survey also concludes with special deep analytical insights based on their opinions. These findings reveal several holes that may be filled using XAI methods, indicating the overall future direction of research in this domain.

In general, this survey intends to provide a comprehensive review of state-of-art XAI applications in the cyber security area. The research motivations behind this work are listed as followings:

(1) To review different techniques and categorizations of XAI.
(2) To review existing challenges and problems of XAI.
(3) To identify the frameworks and available datasets for the XAI-based cyber defensive mechanism.
(4) To review the latest successful XAI-based systems and applications in the cyber security domain.
(5) To identify challenges and research gaps of XAI applications in cyber security.
(6) To identify the key insights and future research directions for applying XAI in the cyber security area.

### B. PREVIOUS SURVEYS

XAI and cyber security have been reviewed mostly separately in previous surveys. However, crossovers have emerged between the two domains. This survey presented a comprehensive introduction of different XAI techniques applied in cyber defensive systems. Our work also provided comprehensive XAI categorizations and analyzed details about the existing challenges and frameworks of XAI for cyber security. Cyber security datasets available for XAI models and the cyber threats faced by XAI models are discussed in this paper as well. Table 1 contrasts our study with currently available surveys and reviewing articles. Many existing surveys only analyzed Artificial Intelligence (AI) applications, either ML or DL, in the cyber security area, whereas other authors review XAI methods for a narrow set of cyber security applications. Some reviewers could not describe the background of XAI and cyber security in detail. Furthermore, most articles discuss only AI applications in cyber security or XAI implemented in other domains rather than focusing on cyber security.

**TABLE 1.** Comparison of existing surveys with our work (legend: √ means included; N/A means not included; ≈ means partially included)

| Survey number | Reference number | Survey year | XAI | | | | | | Cyber security | | | | Key insights and future directions |
|---|---|---|---|---|---|---|---|---|---|---|---|---|---|
| | | | XAI Categorization | XAI Framework | ML | DL | XAI Evaluation | XAI Challenges | Cyber security datasets | Cyber attacks | Industrial applications | Adversarial threats on XAI | |
| 1 | [13] | 2016 | N/A | N/A | √ | N/A | N/A | N/A | N/A | √ | ≈ | N/A | ≈ |
| 2 | [14] | 2016 | N/A | N/A | √ | √ | N/A | N/A | √ | ≈ | ≈ | ≈ | √ |
| 3 | [15] | 2017 | N/A | N/A | N/A | √ | N/A | N/A | √ | √ | ≈ | N/A | √ |
| 4 | [16] | 2018 | N/A | N/A | √ | ≈ | N/A | N/A | √ | √ | ≈ | N/A | √ |
| 5 | [17] | 2018 | N/A | N/A | √ | ≈ | N/A | N/A | √ | √ | ≈ | N/A | √ |
| 6 | [18] | 2019 | N/A | N/A | √ | ≈ | N/A | N/A | √ | √ | ≈ | N/A | √ |
| 7 | [19] | 2019 | √ | √ | ≈ | ≈ | √ | √ | N/A | N/A | √ | N/A | √ |
| 8 | [20] | 2020 | ≈ | ≈ | N/A | N/A | N/A | N/A | √ | N/A | √ | N/A | √ |
| 9 | [7] | 2021 | N/A | N/A | √ | N/A | N/A | N/A | √ | √ | ≈ | N/A | √ |
| 10 | [21] | 2018 | N/A | N/A | √ | √ | N/A | ≈ | ≈ | √ | ≈ | N/A | √ |
| 11 | [22] | 2018 | N/A | N/A | √ | √ | N/A | N/A | √ | ≈ | √ | ≈ | √ |
| 12 | [23] | 2018 | N/A | N/A | √ | √ | N/A | N/A | √ | √ | ≈ | ≈ | ≈ |
| 13 | [24] | 2018 | N/A | N/A | √ | √ | N/A | N/A | √ | √ | ≈ | √ | √ |
| 14 | [25] | 2022 | √ | √ | √ | √ | ≈ | √ | N/A | N/A | √ | N/A | √ |
| 15 | [26] | 2021 | N/A | √ | √ | N/A | ≈ | N/A | √ | √ | N/A | √ | ≈ |
| 16 | [27] | 2021 | √ | √ | ≈ | ≈ | N/A | √ | √ | N/A | N/A | N/A | √ |
| 17 | [28] | 2019 | N/A | N/A | √ | √ | N/A | N/A | √ | √ | ≈ | ≈ | √ |
| 18 | [29] | 2019 | √ | √ | N/A | √ | ≈ | √ | √ | √ | N/A | N/A | √ |
| 19 | [2] | 2019 | √ | √ | ≈ | √ | N/A | √ | √ | √ | N/A | N/A | √ |
| 20 | [9] | 2019 | √ | √ | ≈ | ≈ | ≈ | √ | N/A | √ | ≈ | N/A | √ |
| 21 | [30] | 2022 | √ | N/A | ≈ | ≈ | ≈ | √ | ≈ | √ | √ | √ | √ |
| 22 | [31] | 2021 | ≈ | ≈ | √ | √ | N/A | N/A | √ | √ | ≈ | √ | √ |
| 23 | [32] | 2020 | N/A | N/A | ≈ | ≈ | N/A | N/A | √ | √ | √ | √ | ≈ |
| 24 | [33] | 2020 | √ | √ | √ | N/A | √ | N/A | N/A | √ | N/A | ≈ | √ |
| 25 | [34] | 2021 | ≈ | ≈ | ≈ | N/A | ≈ | N/A | √ | √ | √ | ≈ | √ |
| 26 | [10] | 2021 | √ | √ | √ | √ | ≈ | ≈ | N/A | N/A | N/A | N/A | ≈ |
| 27 | [12] | 2021 | √ | ≈ | √ | √ | N/A | ≈ | N/A | N/A | ≈ | N/A | √ |
| 28 | [35] | 2022 | √ | √ | N/A | N/A | N/A | √ | ≈ | ≈ | √ | N/A | √ |
| 29 | [36] | 2022 | √ | √ | N/A | ≈ | ≈ | N/A | √ | √ | N/A | √ | √ |
| **30** | **Our Paper** | **2022** | √ | √ | √ | √ | √ | √ | √ | √ | √ | √ | √ |

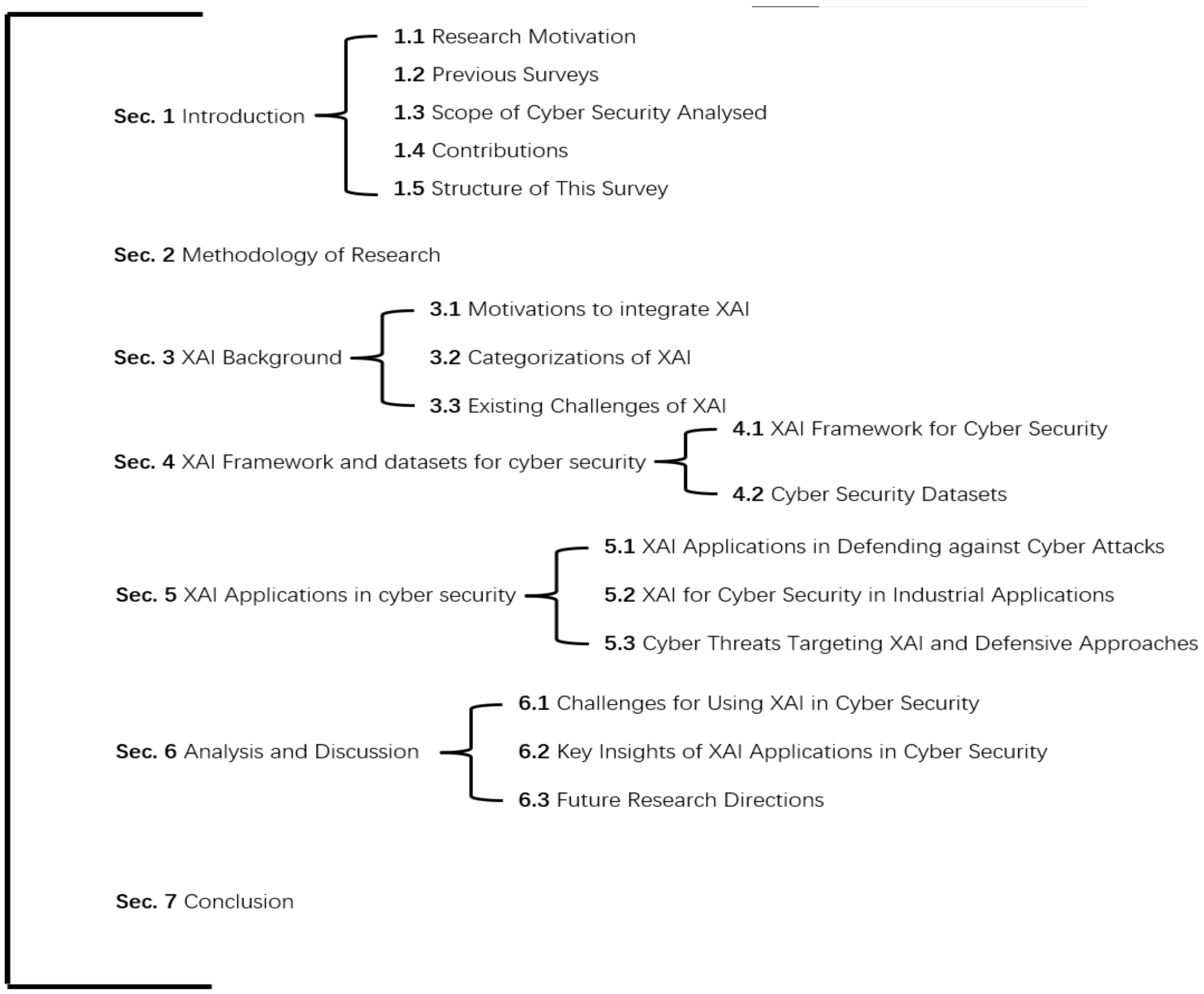

**FIGURE 1. Structure of this paper.**

From Table 1, it is obvious that this survey is comprehensive and distinct in including the following features in comparison to previously published survey research in the field: summarizing commonly used cyber security datasets available, discussing popular XAI tools and their applications in the cyber security area, analyzing the XAI applications in defending different categories of cyber attacks, providing assessment measures for evaluating the performance of XAI models, giving descriptions on the adversarial cyber attacks which XAI itself may suffer, and pointing out some key insights about applying XAI for cyber security.

### *C. SCOPE OF CYBER SECURITY ANALYSED*

In agreement with the International Organization for Standardization (ISO/IEC 27032) [37], cyber security is defined as the privacy, integrity, and availability of internet data. Cyber attacks are cybercriminal attacks undertaken using one or more computers against a single or numerous computers or networks. A cyber assault can purposefully destroy systems, steal data, or utilize a compromised computer as a launch pad for more attacks [38]. Due to the wide spreading of cyber attacks and threats, the cyber security industries are seeing rapid expansion. As a result, by 2026, the worldwide cybersecurity sector is anticipated to be worth 345.4 billion USD [39]. On the other hand, besides the conventional cyber attacks including malware, botnet, and spam, adversarial cyber security threats specifically targeting AI models are Gradually emerging in recent years as well [24]. Therefore, the scope for the domain of cyber security analyzed in this survey paper will be constituted in the following 3 sub-fields in conjunction with XAI:

1) Different categories of the most prominent cyber attacks including malware, Botnet, spam, fraud, phishing, Cyber Physical Systems (CPSs) attacks, network intrusion, Denial-of-service (DoS) attacks, Man-in-the-middle (MITM) attacks, Domain Generation Algorithms (DGAs), and Structured Query Language (SQL) injection attacks are described in detail respectively. By doing so, the terminologies of cyber attacks are clear and the defensive systems against these attacks are discussed in this paper as well.
2) Cyber security implementation in different industrial areas including smart grid, healthcare, smart agriculture, smart transportation, Human-Computer Interaction(HCI), and smart financial system will be reviewed in this survey. This paper provides a brief introduction of XAI for cyber security in each domain respectively.

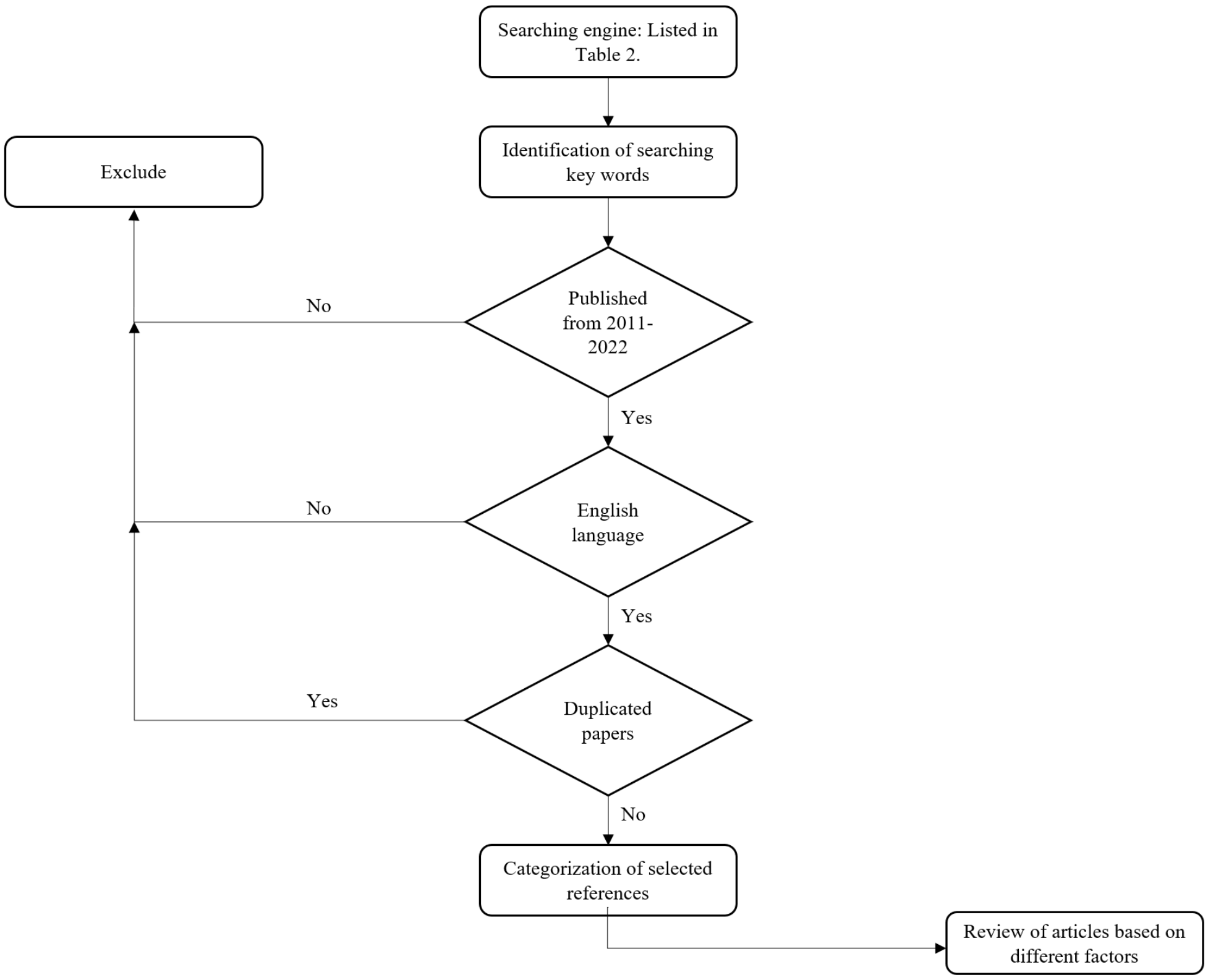

**FIGURE 2.** **Research methodology flow chart.**

3) While XAI is implemented in many different scenarios to defend against cyber threats, XAI models will face adversarial attacks targeting XAI models as well. This survey will investigate cyber security from this perspective as well. Adversarial threats targeting XAI, defense approaches against these attacks, and the establishment of secure XAI cyber systems will be interpreted respectively.

### D. CONTRIBUTIONS

This study extensively evaluates current breakthroughs and state-of-the-art XAI-based solutions in a wide variety of cyber security applications and cyber attack defensive mechanisms to address the gaps and shortcomings mentioned in earlier surveys. There is no previous survey available analyzing the state-of-art XAI applications in cyber security systemically from the perspectives of both cyber attack defensive schemes and industrial applications. Our research's contributions can be summarized in the following points:

1) We rationalize the motivations for integrating XAI in AI-based cyber security models whereas the basic background on XAI is presented.
2) We provide a thorough summary as well as a quick overview of the datasets that are accessible for the usage of XAI applications in cyber security.
3) We discuss different categories of defensive applications of XAI against cyber attacks respectively, and we highlight the advantages and limitations to develop XAI-based cyber-defense systems.
4) We justify XAI for cyber security in different industry scenarios.
5) We illustrate Adversarial cyber threats pointing to XAI models are described whereas the defense approaches against these attacks.
6) We outline the outstanding issues and existing challenges associated with the intersection of XAI and cyber security, and we identify the key insights and future research directions for the XAI applications in cyber security.

### E. STRUCTURE OF THIS SURVEY

As shown in Fig 1, this survey has been organized in such a way that the background information for the research being examined comes first. Section II introduces the methodology of research on this survey in the field of XAI applications in cyber security. Section III discusses the general background of XAI, motivations, categorizations, and challenges of XAI are justified in this section. The section after that (Section IV) is organized based on the XAI framework and available datasets for cyber security. Section V will be devoted to a comprehensive discussion of XAI applications in cyber security from different perspectives. The existing challenges, key insights, and future directions of this area are highlighted in Section VI, which is followed by the conclusion. And the conclusion would be the last section, which is Section VII.

**TABLE 2.** Research searching database engines.

| Searching Engines | Database Address |
|---|---|
| Springer | https://link.springer.com/ |
| Taylor & Francis | https://taylorandfrancis.com/ |
| Semantic Scholar | https://www.semanticscholar.org/ |
| ACM Digital Library | https://dl.acm.org/ |
| ResearchGate | https://www.researchgate.net/ |
| Google Scholar | https://scholar.google.com/ |
| IEEE Xplore | https://ieeexplore.ieee.org |
| Elsevier | https://www.elsevier.com/ |
| Research Rabbit | https://researchrabbitapp.com/ |

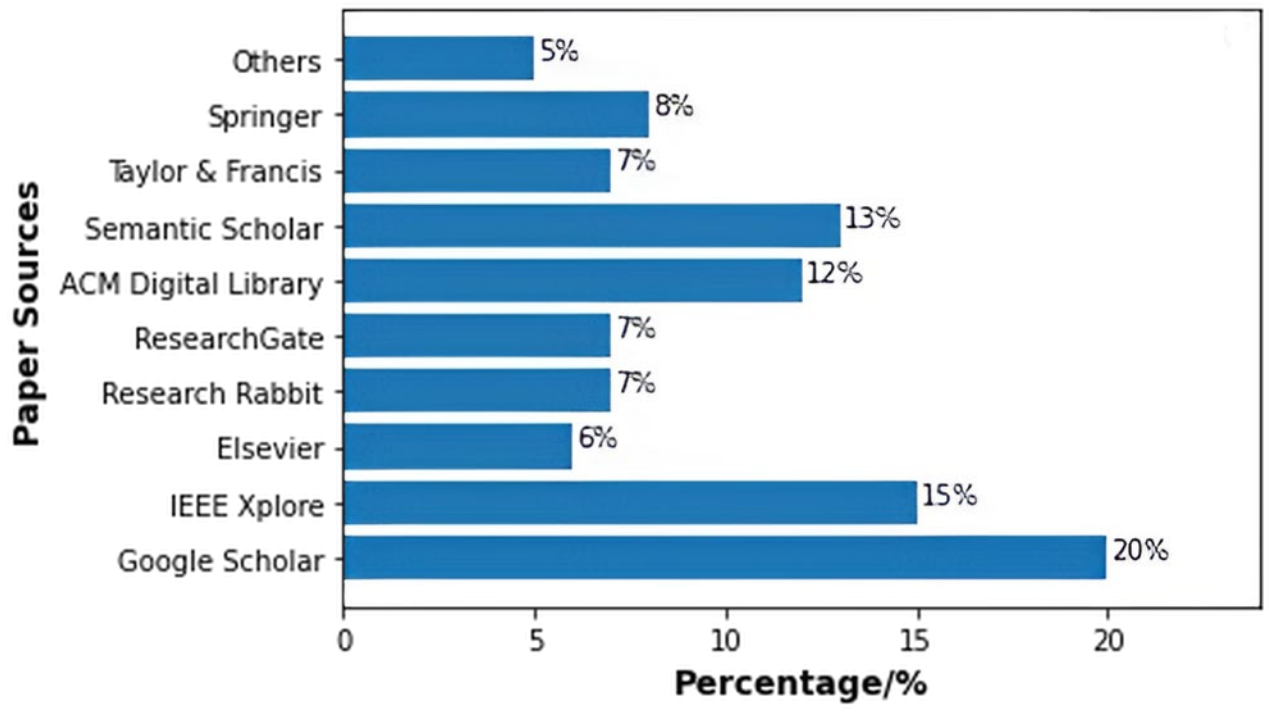

**FIGURE 3.** Percentage of Reviewed Papers from Sources.

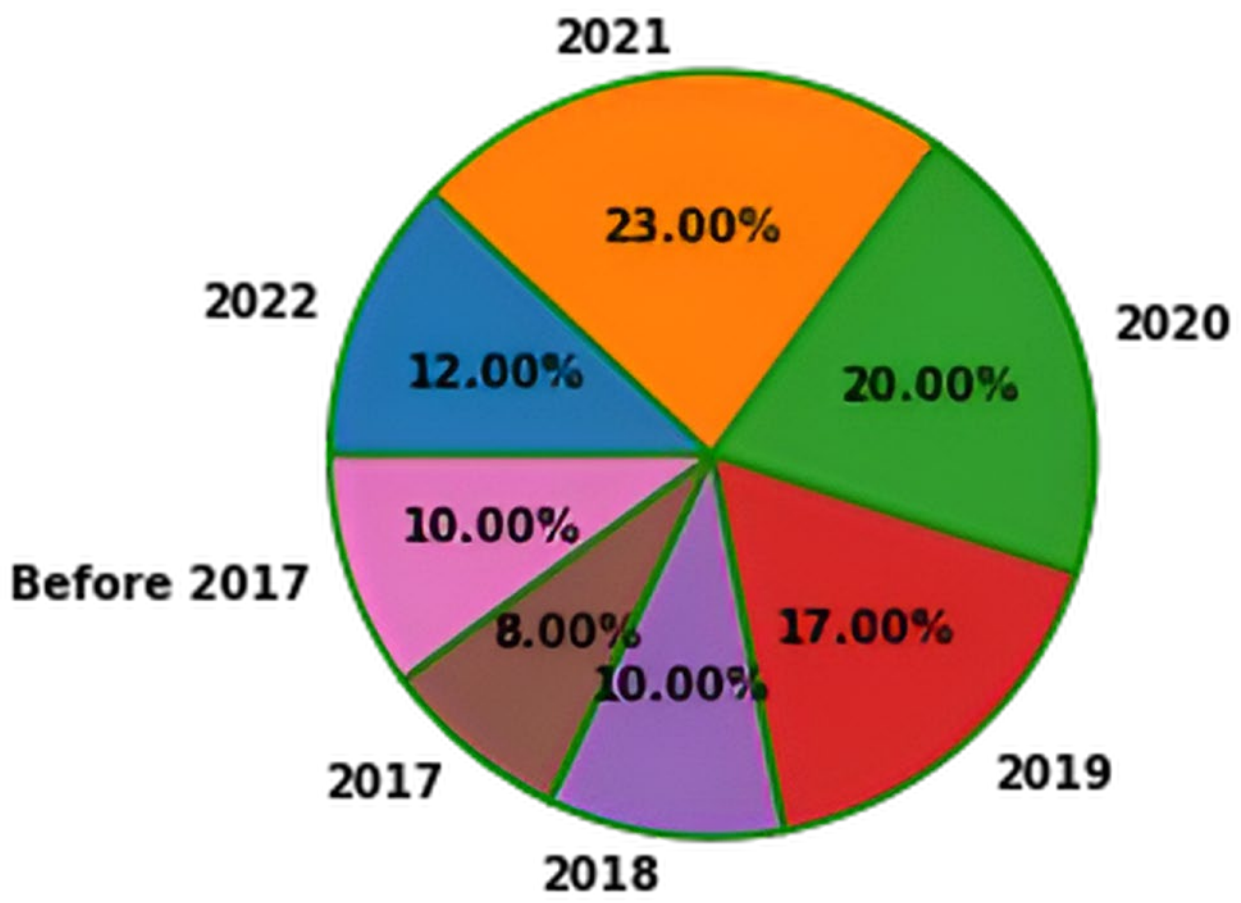

**FIGURE 4.** Percentage of Papers included from 2011 to 2022.

## II. METHODOLOGY OF RESEARCH

The research methodology flow chart of this survey is described in Figure 2. As we mentioned in Section I Introduction, the goal of this study was to investigate the research state-of-art in the areas of XAI applications in cyber security. Therefore, to collect the research articles reviewed, the following criteria were established:

1) A thorough search was carried out whereas different academic search engines illustrated in Table 2 were utilized to collect the relevant papers.
2) The searching keywords for this survey paper were constituted as 2 aspects: “XAI” and “Cyber Security”. To create the search string, all potential pertinent synonyms of the given terms were discovered in different databases and the percentage of reviewed papers from sources was depicted in Figure 3. The following synonyms may be pertinent to the subject: “Cyber Security”, “Cyber Physical”, “Cyber Attack”, “Cyber Threat”, Network Security”, “Cyber Crime”, “XAI”, “Explainable Artificial Intelligence”, “Interpretable Artificial Intelligence”, “Explainable ML (XML)”, and “Transparent Artificial Intelligence”.
3) Only researches published between 2011 and 2022 were selected to report on the most recent trends in the application of XAI techniques in cyber security for this research. Besides, papers published after 2017 were given higher attention and occupied a large proportion of all reviewed publications, as shown in Figure 4.
4) Only publications written in the English language were included in this review and duplicated studies were excluded.
5) Only papers objecting to cyber security vulnerability domains were reviewed in this survey paper whereas researches proposing ML-based systems, DL-based systems, XAI-based mechanisms, and AI-based mechanisms would be extracted.

The procedure of choosing articles was instantaneous and consisted of two steps: firstly, the searching results were initially chosen based on the selection criteria by scanning the publications' titles and abstracts; secondly, the documents chosen in the initial phase were thoroughly read to create a shortlist of articles published that would be chosen based on the inclusion and exclusion criteria.

## III. XAI BACKGROUND

As we introduced in Section I, the concept of XAI is defined as the technique to improve the human understanding of how AI makes decisions [10]. In this section, we will review the general background of XAI, providing some necessary prior knowledge for readers to have a better understanding in the following sections introducing the XAI applications in cyber security.

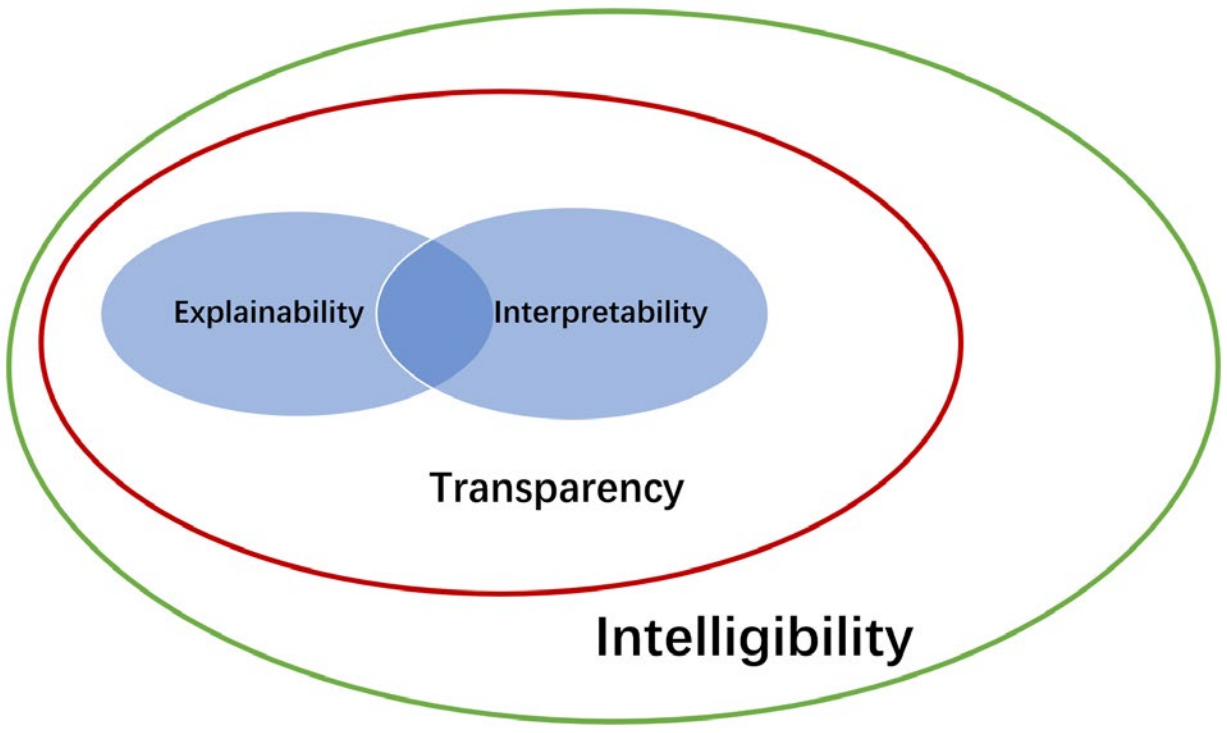

**FIGURE 5** A Venn Diagram showing the connections between words used frequently in the XAI domain.

Before exploring the XAI background deeply, it is worth mentioning and clarifying the terminologies in the XAI domain. Numerous concepts and phrases, which include intelligibility, explainability, transparency, and interpretability. have been used to characterize XAI recently [40]. And the relationships between these terms are shown in Figure 5. Among these terms, interpretability is defined as a concept similar to explainability [41]. However, in recent years, the terminology for the term "interpretability" has shifted to information extraction rather than providing explanations [42], meaning that the terms of interpretability and explainability are becoming more diverse while still intersecting with each other. Therefore, in this study, we focus on the side of "explainability" in XAI whereas the reviewed papers focusing on "intelligibility", "transparency", and "intelligibility" parts would be extracted and excluded according to their clutters with the concept of "explainability".

In the following subsections of this section, we will introduce the background of XAI from different perspectives respectively, including the motivations to integrate XAI into cyber security, categorizations of XAI, and existing challenges of XAI. The purpose of this section is to provide readers with a general description of the XAI area so that readers could have a deeper understanding of the parts of XAI applications in cyber security.

### A. MOTIVATIONS TO INTEGRATE XAI INTO CYBER SECURITY

Given the constant growth in complexity and volume of cyber attacks including malware, intrusion, and spam, coping with them is becoming increasingly difficult [17]. According to [43], conventional algorithms including rule-based algorithms, statistics-based algorithms, and signature-based approaches are utilized to detect intrusions in the cyber security area. However, due to the growing amount of data being communicated over the Internet and the emergency of the new networking paradigms including the Internet of Things (IoT), cloud computing, and fog/edge computing [44], these traditional approaches have a low capacity to process massive amounts of data and high computing costs [7].

On the other hand, Artificial intelligence works as one of the foundational technologies of Industry 4.0 [31]. Therefore, AI techniques including ML algorithms and DL algorithms can play a significant part in the provision of intelligent cyber security services and management in recent years. For instance, Daniele *et al.* [17] concluded the implementation of ML Methods for malware analysis including malware detection, malware similarity analysis, and malware category analysis. And Donghwoon *et al.* [15] utilized DL-based approaches to network anomaly detection and network traffic analysis.

Nevertheless, due to the limitations of the AI-based approaches, the applications of AI in the cyber security area are facing challenges as well. For instance, the access to cybersecurity-related data [45], adversarial attacks on AI models [46], and Ethics and Privacy issues [47] are typical inherent limitations suffered by AI-based cyber security systems. Among these drawbacks, the black-box nature of AI models is a severe limitation that we should pay more attention to when AI models are integrated into the cyber security domain [48]. Because of AI models' black-box characteristics, the cybersecurity-related decisions generated by AI-based models lack rationale and justifiability of their decisions and therefore are difficult for people to understand how these results are produced [49]. In this case, the cyber defensive mechanisms would become black-box systems that are extremely vulnerable to information breaches and AI-based cyber threats [50].

Therefore, to deal with the drawbacks of utilizing AI for cyber security, XAI is a reaction that emerged to the growing black box issue with AI. Users and specialists can understand the logical explanation and main data evidence due to XAI's contribution of interoperability to the results produced by the AI-based statistical models [19].

To conclude, the motivations to apply XAI to cyber security are given as followings:

1) Building trust is a key object for integrating XAI which is closely related to transparency and understanding of cybersecurity-related decision models.
2) Another motivation to apply XAI in the cyber security area is to comply with many new regulations and General Data Protection Regulation (GDPR) laws [51] calling for providing explanations to the entire society in various fields including cyber security.
3) Justice, social responsibility, and risk mitigation are significant concerns for applying XAI in cyber security because protecting cyber security may be dealing with serious social problems, sometimes even human lives, and not just cost-benefit calculations.

4) Cyber security system biases and the misunderstanding of their effectiveness have emerged as key drivers for XAI. For instance, biased training data occurs as a problem that affects the model's output's credibility, in particular when working with neural networks that learn patterns from training data [52].
5) Ability to provide obliged and decent justification for the cyber security system. By doing so, the created cyber security defensive mechanisms can not only be fair and socially responsible for the decisions, but also defend their results with justifications.

## B. CATEGORIZATIONS OF XAI

According to [53], [54], the XAI categories can be structured in a variety of aspects shown in Figure 6. It is noticeable that the categorization methods are not ideal, meaning that overlapping may happen and one specific XAI technique can be categorized into one or more aspects. Therefore, it would be more precise and concrete if we categorized one XAI technique from different categorization perspectives. By doing so, more information and characteristics of this XAI approach could be revealed at different levels.

### 1) INTRINSIC OR POST-HOC

This categorization method distinguishes between achieving explainability by limiting the complexity of the AI model (intrinsic) or by analyzing the methodology of the model after training (Post-hoc) to differentiate whether explainability is achieved. An intrinsic XAI approach produces the explanation concurrently with the forecast by using data that the model emits as a result of the prediction-making process [55]. Some ML models, including Decision Trees and Sparse Linear models, are regarded as intrinsic XAI approaches because they are self-explained. On the other hand, Post-hoc explanations are the utilization of interpretation methods after the models have been trained and the decisions have already been made. Local Interpretable Model-agnostic Explanations (LIME) [56] and Permutation Importance [57] are typical Post-hoc explanation methods working independently as an external interpretable model.

### 2) MODEL-SPECIFIC OR MODEL-AGNOSTIC

XAI methods can also be classified according to the classes of models to that XAI methods could be applied, which are model-specific or model-agnostic. Model-specific explanation tools are specific to a single model or group of models. For instance, the graph neural network explainer [58] is a method for presenting comprehensible justifications for any GNN-based model's predictions on any graph-based ML problem. On the contrary, model-agnostic explanation tools can be implemented with any ML model in theory. Furthermore, model-agnostic explanation methods usually work by analyzing feature inputs and outputs and do not have access to the models' internal information, such as weights or structural information by definition. Shapley Additive Explanations (SHAP) tools [59], Saliency Map [60], and Gradient-weighted Class Activation Mapping (Grad-CAM) [61] are widely used model-agnostic explanation tools.

### 3) LOCAL OR GLOBAL

Explanations of the decision models can be divided as local or global depending on the model's scope. Local explainability describes a system's capacity to show a user why a particular choice or decision was made. Some popular explainability methods such as LIME [56], SHAP [59], and counterfactual explanations [62] can be filed under this category. Local explainability methods are emphasized as the first crucial component of model transparency [55]. In the contrast, global explainability refers to the explanation of the learning algorithm as a whole, taking into account the training data utilized, the algorithms' proper applications, and any cautions regarding the algorithm's flaws and improper applications. Global Attribution Mapping (GAM) is proposed in [63] as a global explanation approach to explain the landscape of neural network predictions across subpopulations.

### 4) EXPLANATION OUTPUT

The explanation output is also a crucial component of XAI categorization for the reason that the format of the explanation output would have a strong influence on certain users. For instance, text-based explanation methods are widely utilized in the field of Natural Language Processing (NLP) to fine-grained information and generate human-readable explanations [64]. On the hand, the visualized explanation approaches are used in vaster domains including NLP [65], neural networks [66], and healthcare [67]. In fact, the majority of feature summary statistics can also be visualized and some feature summaries are only meaningful when visualized [68]. Arguments-based explanations involve outlining the features in a way that humans use to come to decisions to help humans to better understand the relevance of a feature [69]. Model-based explanation approaches need to outline the internal working logic of a black-box model. And this is often accomplished by approximating the black-box model behavior with a different model that is more interpretable and transparent [10]. For instance, Wu *et al.* [70] proposed a model-specific technique aiming to reduce the complexity of the Deep Neural Network (DNN) model by introducing a model complexity penalty function. And Lakkaraju *et al.* [71] proposed a model-agnostic technique called Model Understanding through Subspace Explanations (MUSE), aiming at learning the behavious of a specific black-box model by yielding a small number of tight decision sets.

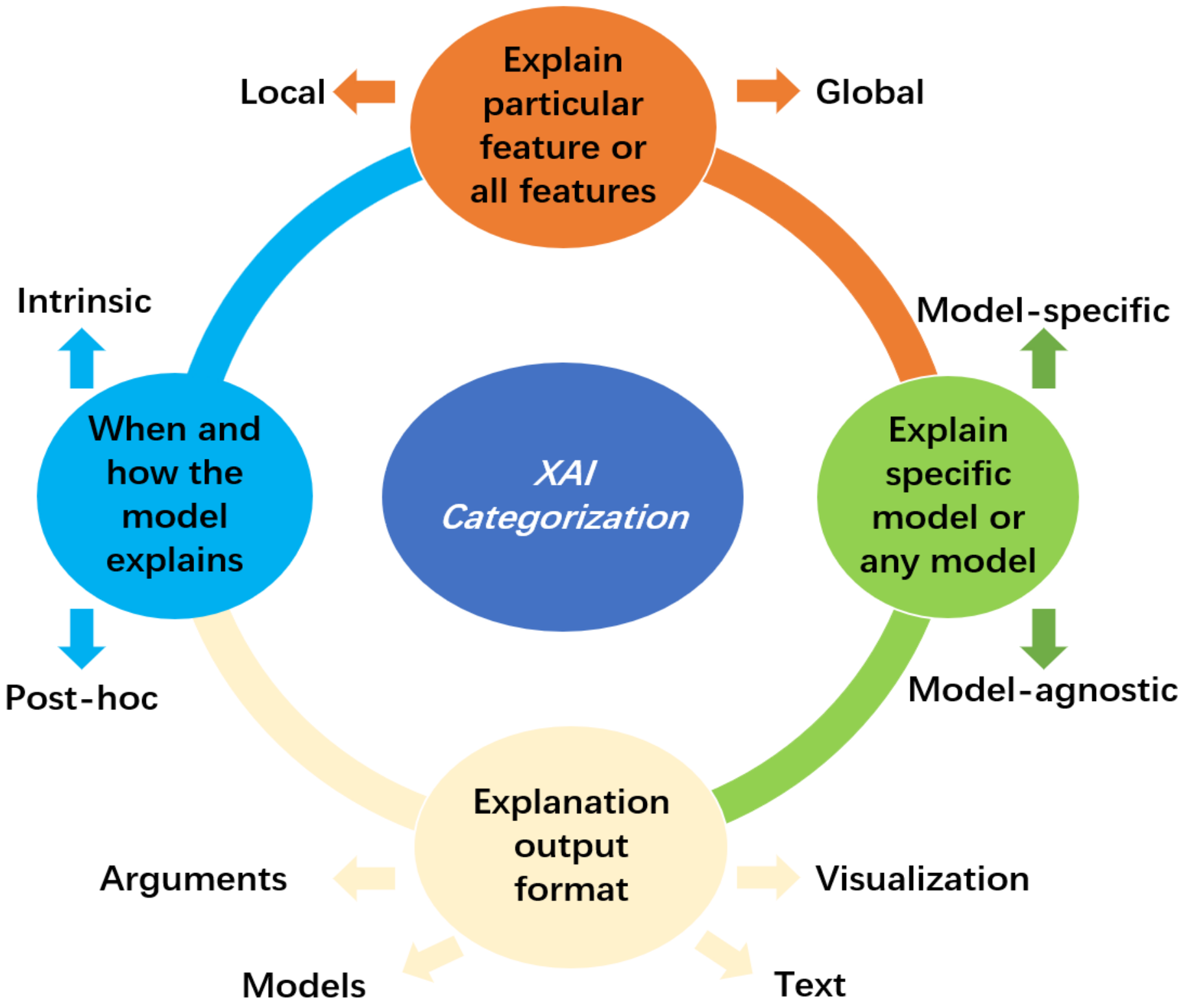

**FIGURE 6** An overview diagram showing the categorization of XAI in different aspects.

### C. EXISTING CHALLENGES OF XAI

Despite the fact that the research community has regarded XAI as a solution to the issues with the trust and dependency posed by conventional black-box AI-based systems, XAI is still facing challenges from different perspectives. Challenges related to XAI security, XAI performance evaluation, legal and privacy issues, and the trade-off between interpretability and accuracy. In Table 3, a summary of challenges related to these challenges of XAI is provided.

#### 1) XAI SECURITY

Some frequently deployed XAI models are susceptible to adversarial attacks, which raises the public's concern about the security of XAI [72].

Guo in [73] highlighted the necessity to develop defense mechanisms that can recognize targeted attacks against XAI engines, especially for the reason that building and quantifying trust between human end-users is essential for 6G to enable higher levels of safety-critical autonomy across a variety of industries. And Fatima *et al.* [74] also pointed out that it would be fascinating to look into the adversarial ML and Deep models (or the application of ML and DL in adversarial circumstances) in XAI and highlighted the three main factors that enable the security of AI models are the changes in the input data used by learning models, bias, and fairness.

Slack *et al.* [75] made criticism about some post-hoc explanation methods such as LIME and SHAP by demonstrating that the extremely biased (racist) classifiers crafted can easily fool these popular explanation techniques. Besides, for the specific Deep Neural Network (DNN) models, Cleverhans *et al.* [76] looked for adversarial vulnerabilities DeepFool tool and offered several methods to harden the model against it.

#### 2) XAI PERFORMANCE EVALUATION

The effectiveness of an XAI method could be evaluated and measured in a variety of ways. However, there is no accepted system available for determining if an XAI system is more user-intelligent than another XAI system at this time [77].

In papers [78] and [79], strong concerns were proposed about choosing the best technique for explainability requires a well-established evaluation system for explainability.

For the evaluation of the explanations given by post-hoc XAI approaches on tabular data, Julian *et al.* [80] proposed a definition of feature relevance in Boolean functions and a testing environment by creating fictitious datasets. And in paper [81], Leila *et al.* solved the issue of the absence of a heatmap quality measurement that is both impartial and widely acknowledged by presenting a framework for evaluating XAI algorithms using ground truth based on the CLEVR visual question answering task.

TABLE 3. Summary of XAI challenges.

| Challenges | Reference | Descriptions |
|---|---|---|
| XAI security | [73] | The necessity to develop defense mechanisms against attacks especially for building 6G industries. |
| | [74] | The application of ML and DL in adversarial circumstances. Be aware of the input data. |
| | [75] | Criticized some post-hoc explanation methods such as LIME and SHAP by fooling these techniques. |
| | [76] | Discussed the DeepFool tool targeting DNN models and offered several methods against it. |
| XAI performance evaluation | [77] | Outlined the fact that there is no accepted system for determining the XAI system's priority. |
| | [78] | Proposed strong concerns about choosing the best technique for explainability |
| | [80] | Proposed a definition of feature relevance in Boolean functions and a testing environment |
| | [81] | Presented a framework for evaluating XAI algorithms based on the CLEVR visual question answering task. |
| Legal and privacy issues | [82] | Proposed concerns about the role of XAI in marketing AI applications. |
| | [83] | The European Commission (EC) has also published ethical guidelines for Trustworthy AI and highlighted privacy. |
| | [84] | GDPR of the EU outlined the human right to contest the decision made and got an explanation of the decision. |
| | [85] | Discussed what degree people have a legal right to an explanation of automated decision-making under EU law |
| The trade-off between interpretability and accuracy | [53] | Outlined the fact that the algorithms that currently perform the best are frequently the least explainable such as DL. |
| | [86] | Pointed out that models' explainability may be compromised in cases when highly engineered or heavy dimensional features are used |
| | [87] | Adopted a multidisciplinary approach to analyze the relevance of explainability for medical AI from different perspectives |
| | [88] | Argued the necessity to apply XAI in clinical practice |

#### 3) LEGAL AND PRIVACY ISSUES

Besides the above described technical challenges, XAI faces significant legal and privacy issues as well. In numerous instances, including some well-known court cases, a history of biased legal and privacy issues was made by XAI systems [89].

Arun [82] proposed concerns about the role of XAI in influencing the privacy calculus of individuals, especially the privacy concerns of customers in marketing AI applications. The European Commission (EC) has also published ethical guidelines for Trustworthy AI as a legal document [83], highlighting the respect for privacy, quality and integrity of data, and access to data.

The General Data Protection Regulation (GDPR) [84] of the EU has added clarification to its information security architecture. In Recital 71, the word ''explanation'' is mentioned, outlining the human right to contest the decision made following such an evaluation and to get an explanation of the decision. Furthermore, Martin [85] investigated whether and to what degree people have a legal right to an explanation of automated decision-making under EU law, particularly when AI systems are involved.

#### 4) THE TRADE-OFF BETWEEN INTERPRETABILITY AND ACCURACY

The Explainability and performance (predictive accuracy) of a model are generally shown to be in trading-off with each other [90]. In fact, there is a demand for explainable models that can attain high performance because the algorithms that currently perform the best are frequently the least explainable (for example, DL) [53].

Despite simple models being frequently favored for their ease of explaining [91], these models' explainability may be compromised in cases when highly engineered or heavy dimensional features are used [86].

Amann *et al.* [87] adopted a multidisciplinary approach to analyze the relevance of explainability for medical AI from different perspectives, showing the necessity to apply XAI in clinical practice even though the primary objective is to give patients the finest care possible [88].

## IV. XAI FRAMEWORK AND DATASETS FOR CYBER SECURITY

### A. XAI FRAMEWORK FOR CYBER SECURITY

In this section, based on the publications we have carefully read in this survey, we provide a general XAI framework diagram for cyber security applications. And the conceptual framework diagram for XAI applications in cyber security is illustrated in Figure 7. This diagram is considered to be as general as it can be to show the processes of applying XAI in the cyber area domains. There are several stages in this workflow whereas certain sample instances are presented in each stage.

The framework workflow starts by determining the types of cyber security tasks, including malware detection, spam detection, and fraud detection, which are defined by the types of cyber attacks facing. The corresponding data such as emails, network traffic, and application activities will be collected and processed in the next stages. Then features representing significant characteristics will be extracted and fed to train different Artificial Intelligence models depending on specific situations. Cyber security test samples will be analyzed and made decisions after the models have been trained. Users can get decisions and explanations explicitly from self-interpretable models whereas the predictions made by black-box modes require explanations of XAI models to make the users requesting for the cyber security tasks satisfied. It is noticeable that this diagram is only a general workflow of XAI applied in cyber security areas, and the details may differ for different tasks specifically.

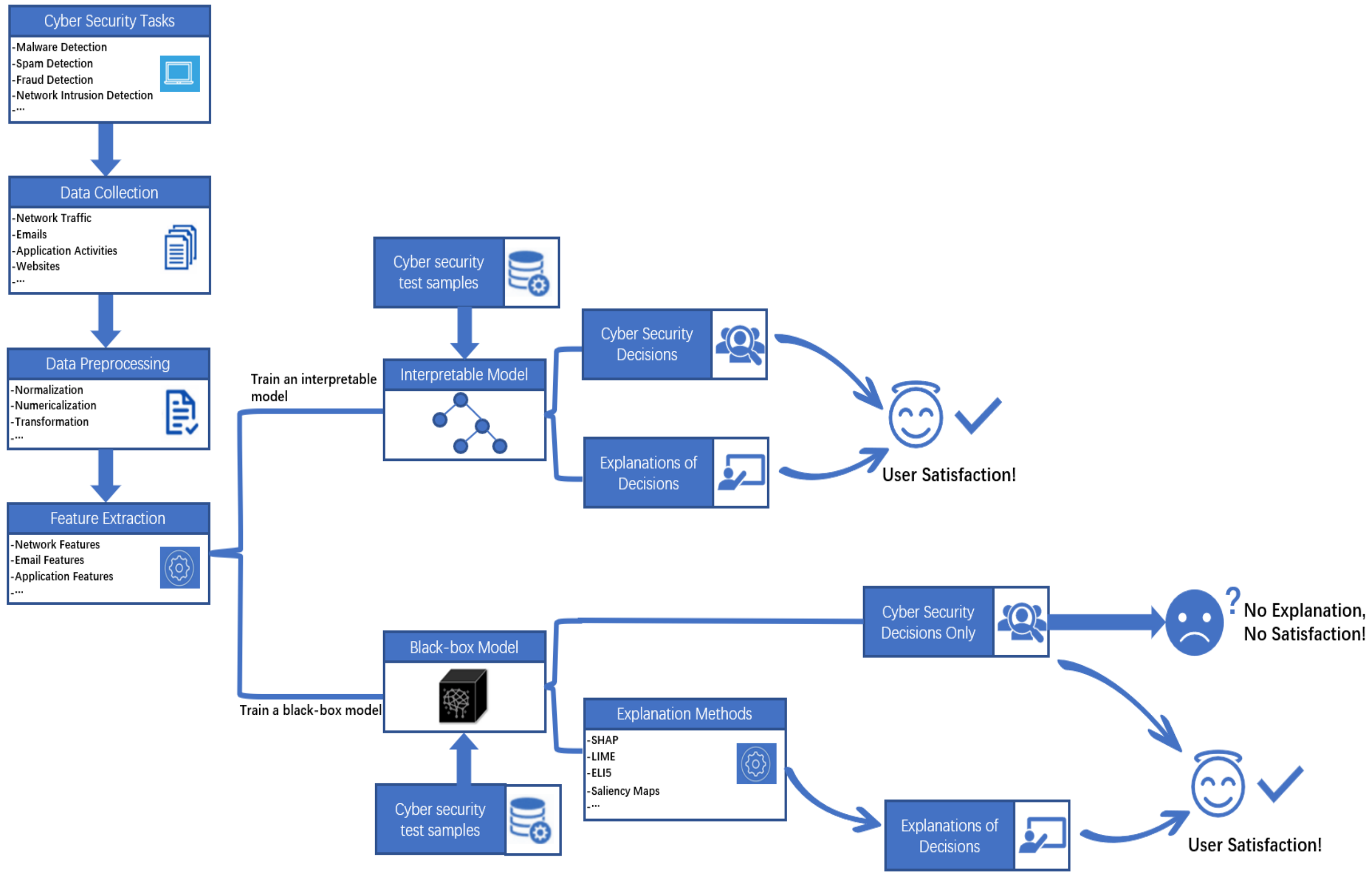

**FIGURE 7** The conceptual framework diagram for XAI applications in cyber security.

## B. CYBER SECURITY DATABASES

It is an undeniable fact that currently judicious selection and use of data is a pretty significant presence for cyber security research [92]. On the other hand, the quality and capacity of data influence significantly the decisions of XAI models, including DL-based models and ML-based models as well. Although cyber security data can be gathered straightforwardly by the use of numerous methods, like using software tools like Win Dump or Wireshark to capture network packets, these methods are mainly targeted and appropriate for gathering narrow or low volumes of data whereas high acquisition time and expenses will be required [93]. Therefore, the utilization of benchmark cyber security datasets can reduce the time spent on data gathering and improve the effectiveness of research. Researchers can train, verify, and evaluate XAI-based cyber security solutions using these benchmark datasets. In this section, we will introduce and describe the most significant datasets employed in cyber security from perspectives of different categories of the most prominent cyber attacks and cyber security implementation in different industrial areas respectively.

Table 4 shows the details of the frequently used public accessible datasets in the context of cyber attacks including malware, Botnet, spam, DGA, DoS, CPSs, phishing, and network intrusion. It is noteworthy that there are some overlappings because some datasets contain several categories of cyber attacks.

On the other hand, Table 5 illustrates a comprehensive overview of XAI applications for cyber security in distinct industries including smart cities, healthcare, smart agriculture, smart transportation, smart financial system, and Human-Computer Interaction(HCI). These industrial datasets can show the potential of applying XAI for cyber security in these domains.

# V. XAI APPLICATIONS TO CYBER SECURITY

This section provides a comprehensive overview of XAI applications in the areas of cyber security from different viewpoints. We categorized these applications into 3 main groups: defensive applications of XAI against cyber attacks, potentials of XAI applications for cyber security in different industries, and cyber adversarial threats targeting XAI applications and defense approaches against these attacks. Some important existing works under each of these domains will be introduced in detail respectively.

**TABLE 4.** Some public available datasets in the context of cyber attacks categories.

| Cyber Attack Categories | Reference | Dataset Name | Year | Cited Number | Dataset Details |
|---|---|---|---|---|---|
| Malware | [94] | N- BaIoT | 2018 | 644 | N-BaIoT contains real traffic (115 numerical features) of 9 commercial IoT devices infected with 2 IoT-based botnets, Mirai and BASHLITE. |
| | [95] | IoTPOT | 2016 | 219 | 500 IoT malware samples from four key families are included in IoTPOT, which was compiled via an IoT honeypot. And these IoT devices were running on different CPU architectures such as ARM, MIPS, and PPC. |
| | [96] | IoT-23 | 2020 | 381 | IoT-23 is a dataset of Internet of Things (IoT) device network traffic. In IoT devices, it has captured 20 malware executions and 3 benign IoT device traffic grabs. |
| | [97] | EMBER | 2018 | 223 | EMBER includes features extracted from 1.1M binary files 200K test samples and 900K training samples (300K harmful, 300K benign, and 300K unlabeled) (100K malicious, 100K benign). |
| | [98] | Genome Project | 2012 | 2689 | More than 1,200 malware samples covering the majority of the current Android malware families were collected in this dataset and were systematically characterized from various aspects. |
| | [99] | VirusShare | Updating | N/A | There are 48,195,237 samples of malware in the collection known as VirusShare. And it is frequently utilized for malware analysis and detection and is primarily affected. |
| | [100] | CICAndMal2017 | 2018 | 143 | Created a new dataset called CI-CAndMal2017 and provide a methodical method to build Android malware datasets using actual smartphones as opposed to emulators. More than 10,854 samples (4,354 malware and 6,500 benign) were collected. |
| | [101] | DREBIN | 2014 | 2102 | DREBIN performs a thorough static analysis of the Android platform to gather as many features of an application as feasible. 5,560 applications from 179 different malware families were collected. |
| Spam | [102] | SMS Spam v.1 | 2011 | 367 | This dataset offered a new real, public, and non-encoded SMS spam collection. |
| | [103] | EnronSpam | 2006 | 743 | The Enron Corpus is a database of over 600,000 emails generated by 158 employees of the Enron Corporation. |
| | [104] | ISCX-URL2016 | 2016 | 100 | Around 114,400 URLs were collected initially in this dataset containing benign and malicious URLs in four categories: Spam, Malware, Phishing, and Defacement. |
| Network Intrusion | [105] | NSL-KDD | 2009 | 3730 | To solve the issues of the KDD data set, a new data set, NSL-KDD, is proposed, which consists of selected records of the complete KDD data set. |
| | [106] | UNB ISCX 2012 | 2012 | 1027 | The Canadian Institute for Cybersecurity at the University of New Brunswick (UNB) established UNB ISCX 2012 in 2012. Over seven days, traffic was recorded in a simulated network environment. |
| | [107] | AWID | 2016 | 365 | A labeled dataset with an emphasis on 802.11 networks is called AWID [117. To collect WLAN traffic in a packet-based format, a small network environment with 10 clients was created, and 15 distinct attacks were carried out. |
| | [108] | CIC-IDS2017 | 2018 | 1672 | The CIC-IDS2017 dataset includes a variety of user-profiles (creating background traffic) and multistage attacks such as Heartbleed and DDoS. Eighty traffic features were extracted using the CICFlowMeter program. |
| | [109] | CIC-DDoS2019 | 2019 | 309 | The CIC-DDoS2019 dataset contains a wide variety of DDoS assaults that were executed utilizing TCP/UDP application layer protocols. |
| | [110] | TON_IoT | 2020 | 103 | TON IoT dataset was constituted by the IoT traffic collected from a medium-scale network at the Cyber Range and IoT Labs of the UNSW Canberra, Australia. Other types of IoT data include operating system logs and telemetry data. |
| | [111] | LITNET-2020 | 2020 | 44 | Feature vectors produced during 12 assaults on common computers installed on an academic network are included in the LITNET-2020 dataset. |
| | [112] | ADFA-LD | 2013 | 281 | The ADFA-LD12 represents a worthy successor to the KDD collection. The most recent publicly accessible exploits and techniques are used with a contemporary Linux operating system for this new dataset. |
| | [113] | UNSW-NB15 | 2015 | 1419 | This dataset contains two label attributes: the first label specifies the attack, while the second label is binary. It also has 49 characteristics. This dataset takes into account assaults such as worms, backdoors, shellcode, DoS assaults, generic assaults, exploits, and analysis assaults. |
| Botnet | [114] | CTU-13 | 2014 | 606 | Raw pcap files for malicious, typical, and background data are included in the CTU-13 dataset. In this dataset, the unidentified traffic is coming from a sizable network, the botnet attacks are real, meaning that it is not a simulated dataset. |
| | [108] | CIC-IDS2017 | 2018 | 1672 | The CIC-IDS2017 dataset includes a variety of user-profiles (creating background traffic) and multistage attacks such as Heartbleed and DDoS. Eighty traffic features were extracted using the CICFlowMeter program. |
| | [115] | ISOT Botnet Dataset | 2011 | 325 | The ISOT HTTP botnet dataset consists of two traffic captures malignant DNS information for nine different botnets and benign DNS information for 19 different well-known software programs. And the ISOT dataset is the combination of several existing publicly available malicious and non-malicious datasets. |
| | [116] | BOT-IOT Dataset | 2019 | 526 | The proposed BOT-IOT Dataset is made up of three parts: network platforms, fictitious IoT services, and features extraction and forensic analytics. |
| | [98] | Genome Project | 2012 | 2689 | More than 1,200 malware samples covering the majority of the current Android malware families were collected in this dataset and were systematically characterized from various aspects. |
| DGA | [117] | UMUDGA | 2020 | 25 | Proposed a comprehensive, labeled dataset with over 30 million AGDs arranged into 50 groups of malware variants that are ready for ML. |
| | [118] | AmritaDGA | 2019 | 16 | AmritaDGA is made up of two data sets. The first data collection is gathered from sources that are openly accessible. The second set of information is gathered from a private real-time network. |
| Phishing | [104] | ISCX-URL2016 | 2016 | 100 | Around 114,400 URLs were collected initially in this dataset containing benign and malicious URLs in four categories: Spam, Malware, Phishing, and Defacement. |
| CPSs | [119] | HAI Dataset 1.0 | 2020 | 25 | The HAI dataset was collected from a realistic industrial control system (ICS) testbed augmented with a Hardware-In-the-Loop (HIL) simulator that emulates steam-turbine power generation and pumped-storage hydropower generation. |
| | [120] | Power System Attack Datasets | 2014 | 248 | This dataset consists of three datasets that measure the normal, disturbed, controlled, and cyberattack behaviors of the electric transmission system. The collection contains measurements from relays, a simulated control panel, synchrophasor measurements, and data logs from Snort. |
| DoS | [121] | InSDN Dataset | 2020 | 50 | A variety of attack types, including DoS, DDoS, Web, Password-Guessing, and Botnets, are included in the InSDN dataset. |
| | [106] | UNB ISCX 2012 | 2012 | 1027 | The Canadian Institute for Cybersecurity at the University of New Brunswick (UNB) established UNB ISCX 2012 in 2012. Over seven days, traffic was recorded in a simulated network environment. |

**TABLE 5.** Some public available datasets in the context of distinct industries.

| Different Industry Verticals | Reference | Dataset Name | Year | Cited Number | Dataset Details |
|---|---|---|---|---|---|
| Healthcare | [122] | PPMI | 2011 | 1059 | The PPMI dataset will include 200 healthy volunteers and 400 recently diagnosed PD patients who will be followed longitudinally for clinical, imaging, and biospecimen biomarker assessment at 21 clinical sites utilizing standardized data gathering techniques. |
| | [123] | CoAID | 2020 | 133 | This dataset included bogus news on websites and social media platforms, as well as consumers' social engagement with such material. CoAID (Covid-19 heAlthcare mIsinformation Dataset) featured a variety of COVID-19 healthcare misinformation. CoAID has 4,251 news items, 296,000 user interactions, 926 posts on social media sites regarding COVID-19, and ground truth labels. |
| | [124] | Heart Disease Cleveland UCI | 2020 | 27 | The Heart Disease Cleveland UC Irvine dataset uses 13 factors to predict whether or not a person has heart disease. Reprocessing was done using the 76 feature original dataset. |
| | [125] | MIMIC-III | 2016 | 4140 | MIMIC-III ('Medical Information Mart for Intensive Care') is a sizable, single-center database that contains data on people who have been admitted to tertiary care hospitals' critical care units. |
| | [126] | MIMIC-II | 2011 | 1104 | There were 25,328 stays in intensive care units in the MIMIC-II database. Laboratory data, therapeutic intervention profiles like vasoactive medication drip rates and ventilator settings, nursing progress notes, discharge summaries, radiology reports, and provider order entry data were all collected by the researchers during their detailed examination of intensive care unit patient stays. |
| | [127] | PTB-XL | 2020 | 171 | This 10-second-long 12-lead ECG-waveform dataset has 21837 records from 18885 patients. Up to two cardiologists annotated the ECG waveform data as a multi-label dataset with diagnostic labels further grouped into super and subclasses. |
| | [128] | BreakHis | 2016 | 725 | BreakHis was composed of 9,109 microscopic images of breast tumor tissue collected from 82 patients using different magnifying factors (40X, 100X, 200X, and 400X). To date, it contains 2,480 benign and 5,429 malignant samples |
| | [129] | CPSC2018 | 2018 | 204 | One normal ECG type and eight abnormal ECG types are part of the data utilized in dataset CPSC2018. This study describes the data source, recording details, and clinical baseline characteristics of patients, such as age, gender, and so on. It also describes the typical procedures for detecting and categorizing the aberrant ECG patterns mentioned above. |
| | [130] | REMBRANDT | 2018 | 90 | The 671 cases in the Rembrandt brain cancer dataset were gathered from 14 collaborating institutions between 2004 and 2006. It is available for use with the Georgetown Database of Cancer (G-DOC) open access platform for undertaking clinical translational research. |
| | [131] | GlioVis | 2016 | 446 | GlioVis contains over 6500 tumor samples of approximately 50 expression datasets of a large collection of brain tumor entities (mostly gliomas), both adult and pediatric. |
| Smart Transportation | [132] | Cologne Vehicular mobility trace | 2013 | 327 | During 700.000 individual car excursions are included in the resultant synthetic trace of the car traffic in the city of Cologne, which spans a 400 square kilometer area over the course of a normal working day. |
| | [133] | PKLot | 2015 | 227 | 695,899 photos from two parking lots were collected for this new parking lot dataset using three different camera perspectives. The acquisition methodology enables the collection of static photographs illustrating variations in illumination on sunny, cloudy, and wet days. |
| | [134] | PEMS-SF Data Set | 2011 | 362 | This dataset describes the various car lanes on the motorways in the San Francisco Bay area's occupancy rate, which ranges from 0 to 1. Every ten minutes, samples are taken from the measurements, which span the period from January 1st 2008 to March 30th 2009. |
| | [135] | CNRPark+EXT | 2017 | 282 | The CNRPark+EXT dataset, which was created on a parking lot with 164 spaces, has around 150,000 annotated pictures (patches) of vacant and occupied parking places. |
| | [136] | VED | 2020 | 24 | This open dataset records the GPS positions of moving objects combined with time-series data on their consumption of fuel, energy, speed, and auxiliary power. Between November 2017 and November 2018, a diversified fleet of 264 gasoline vehicles, 92 HEVs, and 27 PHEV/EVs were on the road. The data were gathered using onboard OBD-II recorders. The types of driving situations and seasons range from highways to congested city areas. |
| Smart Cities | [137] | T-Drive | 2011 | 826 | The dataset tracks 10357 taxi movements in Beijing over the course of one week, from February 2 to February 8, 2008. Using longitude and latitude, this data displays the location of a cab continuously throughout a range of time periods. |
| | [138] | GeoLife GPS Trajectories | 2009 | 2328 | The dataset captured a trajectory position that tracks 182 mobile users in Beijing, China, over the course of three years, from April 2007 to October 2011. Over 48,000 hours and nearly 1.2 million kilometers are covered throughout the complete journey. |
| | [139] | KITTI | 2013 | 5831 | A cutting-edge dataset obtained from a Volkswagen station wagon for use in studies on mobile robotics and autonomous driving. a range of sensor modalities, including high-resolution color and grayscale stereo cameras, a Velodyne 3D laser scanner, and a high-precision GPS/IMU inertial navigation system, were used to record 6 hours' worth of traffic scenarios at 10-100 Hz in total. |
| Smart Agriculture | [140] | Images on plant health | 2015 | 550 | Through the current web platform PlantVillage, this dataset made available over 50,000 highly curated photos of healthy and diseased leaves of crop plants. |
| | [141] | PS-Plant | 2019 | 36 | Presented PS-Plant, a low-cost and portable 3D plant phenotyping platform based on an imaging technique novel to plant phenotyping called photometric stereo (PS). |
| | [142] | Plant Pathology | 2020 | 14 | 3,651 high-quality, realistic photos showing the symptoms of various apple foliar diseases were recorded in this collection, together with variations in noise, illumination, angles, and surfaces. The Kaggle community was given access to a subset that had been expertly annotated to provide a prototype dataset for apple scab, cedar apple rust, and healthy leaves. |
| HCI | [143] | Clarkson | 2015 | 73 | This dataset offered a brand-new keystroke dataset that includes 39 users' transcribed text, free text, and short sentences. This dataset can be used to recreate the authentication performance that was seen in earlier studies. However, all participants are required to complete the same set of predetermined activities in a university lab using the same HTML form and desk-top computer. |
| | [144] | Torino | 2005 | 607 | Although the Orino dataset is similarly gathered using a predefined HTML form, participants are free to use any keyboard and complete their tasks at home rather than in a lab. |
| | [145] | Buffalo | 2016 | 51 | This dataset included unprocessed keystroke data from 157 participants who were permitted to freely transcribe fixed text and respond to questions. The dataset is designed to capture the temporal changes in typing habits as well as the disruptions brought on by various keyboard layouts. |
| Smart Financial System | [146] | Nielsen Dataset | 2017 | 32 | This information was gathered between 2006 and 2010 at 35,000 participating mass merchandisers, pharmacies, and grocery stores spread over 55 MSA (metropolitan statistical areas) in the United States. |
| | [147] | Statlog (German Credit Data) Data Set | 1994 | N/A | The German Credit Data provides information on 20 criteria and classification of 1000 loan applicants as either Good or Bad Credit Risks. Also comes with a cost matrix. |

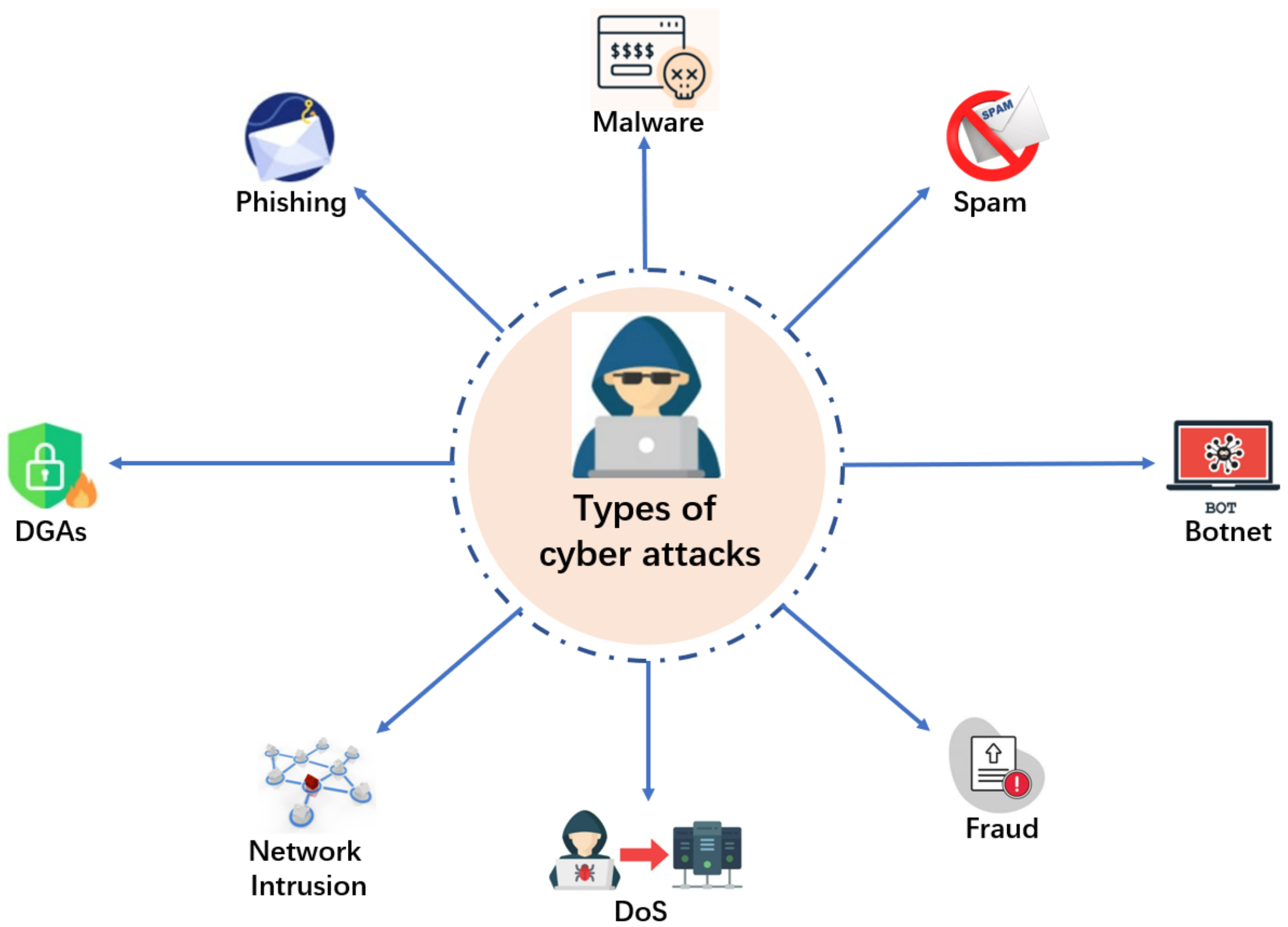

**FIGURE 8** The overview of some common types of cyber attacks.

## A. XAI APPLICATIONS IN DEFENDING AGAINST CYBER ATTACKS

XAI is playing an increasingly significant role in fighting a wide range of cyber attacks, as shown in Figure 8. In this subsection, we analyzed the state-of-art XAI-based defense systems for different categories of cyber attacks. And the conjunctions of these systems with XAI topologies are shown in Table 6 as well.

### 1) MALWARE

One of the major cyber security risks on the Internet today is malware, and implementing effective defensive measures necessitates the quick analysis of an ever-growing volume of malware quantities [148]. Existing techniques for malware detection can be categorized into two main types: Static detection and Dynamic detection [149]. Static malware detection analyzes the malware binary without actually running the code. Instead, the decompilation tool is utilized to obtain the decompiled codes and the included instructions are inspected. However, this kind of strategy can be easily countered by using evading methods like obscuring and incorporating syntax flaws. On the other hand, dynamic malware detection entails executing the malware codes on the testing system and monitoring how it behaves.

In practice, using these conventional malware detection techniques and manually analyzing every malware file in an application takes a lot of time and resources. Therefore, many AI-based malware detection systems, especially DL algorithms are utilized to detect malware with higher better performance and fewer resources than traditional malware detecting methods [150]. However, the working functions of neural networks are similar to a black box, and this topology offers no indication of how it operates [151]. Due to similar motivations, many researchers deploy different categories of XAI approaches in different degrees to make the AI-based malware detection systems more explainable and transparent so that a reliable malware detector can continue to perform well when deployed to a new environment.

There are multiple ways to explain the malware detector. Identifying the most significant local features can always provide valuable explanations for malware detection decisions. Marco *et al.* [152] implemented a gradient-based approach to identify the most influential features contributing to each decision. A popular Android malware detector named Drebin [153] extracted the information from the Android applications. The explainabilities of Drebin on non-linear algorithms, including Support Vector Machines (SVMs) and Random Forests (RFs) are retained by both local explanations and global explanations. The top 10 important features, sorted by their applicability values are disclosed for 3 different cases whereas the AUC remains above 0.96.

For neural network-based detecting mechanisms, Shamik *et al.* [154] proposed a framework explaining how a deep neural network generalizes real-world testing set in different layers. The gradients and weights of different layers of the MalConv architecture [155] and emberMalConv [156] are analyzed to identify different parts' contributions to the classification. High gradient values were found in the header of the files while there are peaks elsewhere, demonstrating that these parts are mostly responsible for classification results. Besides, two filters A and B learned two different sets of features, the accuracy and F1-Score can achieve 91.2% and 90.7% respectively when model B was replaced by model A.

Hamad *et al.* [157] developed a pre-trained Inception-v3 CNN-based transfer learned model to analyze malware in IoT devices. To better understand the features learned by the CNN models, Gradient weighted class activation mapping (Grad-CAM) is utilized to generate cumulative heatmaps and explain the models visually. Besides, t-distributed stochastic neighbor embedding (t-SNE) is used to verify the density of the features in the proposed CNN models. Achieved by the suggested methods, the detection accuracies were 98.5% and 96.9% on the available testing dataset with SoftMax classifier and RF classifier respectively.

Anli *et al.* [158] suggested a technique for extracting rules from a deep neural network so that the rules can be used to identify mobile malware behaviors. To represent the rules discovered between the inputs and outputs of each hidden layer in the deep neural network, an input-hidden tree and a single hidden-output tree for each hidden layer were established. Then the hidden-output tree can tell the most important hidden layer which could specify the related input-hidden tree. The experimental results illustrated accuracy, precision, recall, and F-Measure of the proposed method were 98.55%, 97.93%, 98.27%, and 98.04% respectively.

Giacomo *et al.* [159] offered a way for assessing DL models for malware classification using image data. It uses data from a Grad-CAM and makes an effort to extend the evaluation of the training phase of the models being studied and provide visual information to security analysts. Besides, this technique extends the use of the Grad-CAM and, in addition to the cumulative heatmap, automates the analysis of the heatmaps, assisting security analysts in debugging the model without having any prior knowledge of the issue/pattern in question. Over a testing dataset of more than 8,000 samples classified into 7 families, the proposed model tested in the experimental study had a test accuracy of 97%. However, the limitation of this approach is the morphed version of the malicious sample belonging to the family can evade antimalware detection.

TrafficAV, an effective and explainable detection framework of mobile malware behavior using network traffic was proposed by Shanshan *et al.* [160]. This framework provided explainability to users by defining four sets for each feature extracted from the malware HTTP request and every decision would be distributed certain values to each set respectively, showing the contribution of different sets of features to the detection results. The detection rates of TCP flow and HTTP models reach 98.16% and 99.65% while the false positive rates are 5.14% and 1.84%.

An explainable fast, and accurate approach for detecting Android malware called PAIRED was illustrated by Mohammed *et al.* in [161]. The proposed detection system achieved lightweight by reducing the number of features by a factor of 84% and deploying classifiers that are not resource-intensive. 35 static features were extracted and explained later by SHAP methods. In the experiment, PAIRED malware detection system was able to retain a very high accuracy of 97.98% while processing data in just 0.8206μs by testing with the CICMalDroid2020 dataset with the extracted 35 features.

Martin *et al.* [162] presented a novel way to find locations in an Android app's opcode sequence that the CNN model considered crucial and that might help with malware detection. CNN was demonstrated to assign a high priority in locations similar to those highlighted by LIME as the state-of-the-art for highlighting feature relevance on the benchmark Drebin [101] dataset. And satisfying experimental results were produced as well, including accuracy = 0.98, precision =0.98, recall = 0.98, and F1-Score = 0.97.

### 2) SPAM

Due to the increasing number of Internet users, spam has become a major problem for Internet users in recent years [163]. According to [164], while over 306.4 billion emails were sent and received per day in 2021, spam emails accounted for more than 55 percent of all emails sent in 2021, meaning that unsolicited email messages accounted for nearly half of all email traffic.

Recently, AI-based systems can be regarded as an efficient option to tackle the spam issue primarily because of their ability to evolve and tune themselves [165]. However, due to the privacy and legal specialties of spam, users can ask many questions about AI models, especially the black-box ML and DL models [166]. For instance, a curious spam recipient can have an interest in understanding the utilized AI models and ask the following questions:

1) Why is Message classified as spam by Model?
2) What distinguishes spam from no spam?
3) How does Model distinguish spam from no spam?
4) How does Model work distinguishing an alternative spam filter Model′ used in the past?
5) How does Model work?

These proposed questions can be answered by the implementation of XAI algorithms and XAI algorithms can be used to complement ML models with desired properties, such as explainability and transparency [167]. And many works of literature have studied this area to enhance the trust of the AI-based spam filters.

Julio *et al.* [168] conducted a highly exploratory investigation on fake spam news detection with ML algorithms from a large and diverse set of features. SHAP method was deployed to explain why some are classified as fake news whereas others are not by representative models of each cluster. Novel features related to the source domain of the fake news are proposed and demonstrated five times more frequencies appeared in the detection models than in other features. Besides, only 2.2 percent of the models have a detection performance higher than 0.85 in terms of AUC, which highlights how difficult it is to identify bogus news.

The legally required trade-off between accuracy and explainability was discussed and demonstrated in the context of spam classification by Philipp *et al.* in [169] as well. A dataset of 5574 SMS messages [170] was used to support the argument that it is equally important to select the appropriate model for the task at hand in addition to concentrating on making complex models understandable. In this work, under circumstances, that which just a small quantity of annotated training data is available, very simple models, such as Naive Bayes, can outperform more complicated models, such as Random Forests.

HateXplain, a benchmark dataset for hate speech spam that considers bias and explainability from many angles was introduced by Binny *et al.* in [171]. Several models including CNN-GRU [172], BiRNN [173], and BiRNN-Attention [174] were used and tested on this dataset whereas explainability-based metrics such as Intersection-Over-Union (IOU), comprehensiveness, and sufficiency were utilized to evaluate the model interpretability. Experimental results showed that models that succeed at classification may not always be able to explain their conclusions in a way that is believable and accurate. The limitations behind this benchmark dataset are that external contexts that would be relevant to the classification task, such as the profile bio, user gender, and post history were not considered and the proposed dataset contained English language only.

#### 3) BOTNET

A botnet attack is known as a group of connected computers working together to carry out harmful and repetitive actions to corrupt and disrupt the resources of a victim, such as crashing websites [175]. As shown in Figure 9, a typical botnet's lifecycle contains 5 phases, including Initial Injection, Secondary Injection, Connection, Malicious Activities, and Maintenance and Updating.

The market for global botnet detection is anticipated to expand from US$207.4 million in 2020 to US$965.6 million in 2027, at a compound annual growth rate (CAGR) of 24.0 percent from 2021 to 2027, according to [176]. And Imperva Research Labs [177] also found that botnets constituted 57% of all attacks against e-commerce websites in 2021. These statistics indicate that developing AI-based systems for detecting botnets is necessary. Besides, XAI can contribute to the botnet detecting systems' trust and prevent automation bias when users have too much trust in the systems' output.

In [178], HATMA *et al.* proposed a novel model for botnet DGA detection. Five ML algorithms were utilized and tested with datasets of 55 botnet families. Random Forest achieved the best accuracy of 96.3% and outperformed previous works as well. Open-source intelligence (OSINT) and XAI techniques including SHAP and LIME were combined in this work to provide an antidote for skepticism toward the model's output and enhance the system trust. Besides, the limitations of the proposed frameworks were the temporal complexity involved in calculating the characteristics and the model's low resistance to Mask botnet assaults.

Shohei *et al.* [179] presented a novel two-step clustering approach based on DBSCAN to cluster botnets and classify their categories. Important features were represented and explained by combining subspace clustering and frequent pattern mining from 2 different real-world flow datasets: MAWI [180] and ISP. 60 bot groups from 61,167 IP addresses were categorized from the MAWI dataset whereas 295 bot groups from 408,118 IP addresses from the ISP dataset. And the cluster results of botnets were self-explained by using a dendrogram.

Visualization tools are also used to give better explanations about the reasons for labeling an account as botnet or legitimate. Michele *et al.* [181] suggested ReTweet-Tweet (RTT), a small but informative scatterplot representation to make it simpler to explore a user's retweeting activities. While the proposed botnet detection method Retweet-Buster (RTbust) based on Variational autoencoders (VAEs) and long short-term memory (LSTM) network unsupervised feature extraction approaches were utilized in a black-box nature, the visualization tool RTT can still be employed economically after RTbust has been applied to comprehend the traits of those accounts that have been classified as bots.

Some researchers suggested the necessity to reduce the number of the required features for botnet classification to overcome the scalability and computation resource problems and provide more reliable explanations in botnet detection systems. In [182], Hayretdin *et al.* utilized Principal Component Analysis (PCA) for feature dimension reduction Decision Tree classifier preserved the original features and clearly illustrated how the classifier determined the labels. Therefore, An analyst for cyber security can quickly comprehend an attack or typical behavior and utilize this understanding to further interpret a security event or incident.

With the rise of DL (DL), several pilot studies have been created to understand the behavior of botnet traffic. However, It is difficult for users to understand and put their trust in the outcomes of present DL models because of neural networks' poor decision-making and lack of transparency compared to other approaches. To address this issue, Partha *et al.* [183] carried out in-depth tests using both synthetic and actual network traffic produced by the IXIA BreakingPoint System and the results showed that the proposed DCNN botnet detection models outperformed the existing ML models with an improvement of up to 15% for all performance metrics while SHAP was deployed to provide a clear explanation of the model decision and gain the trust of the end users.

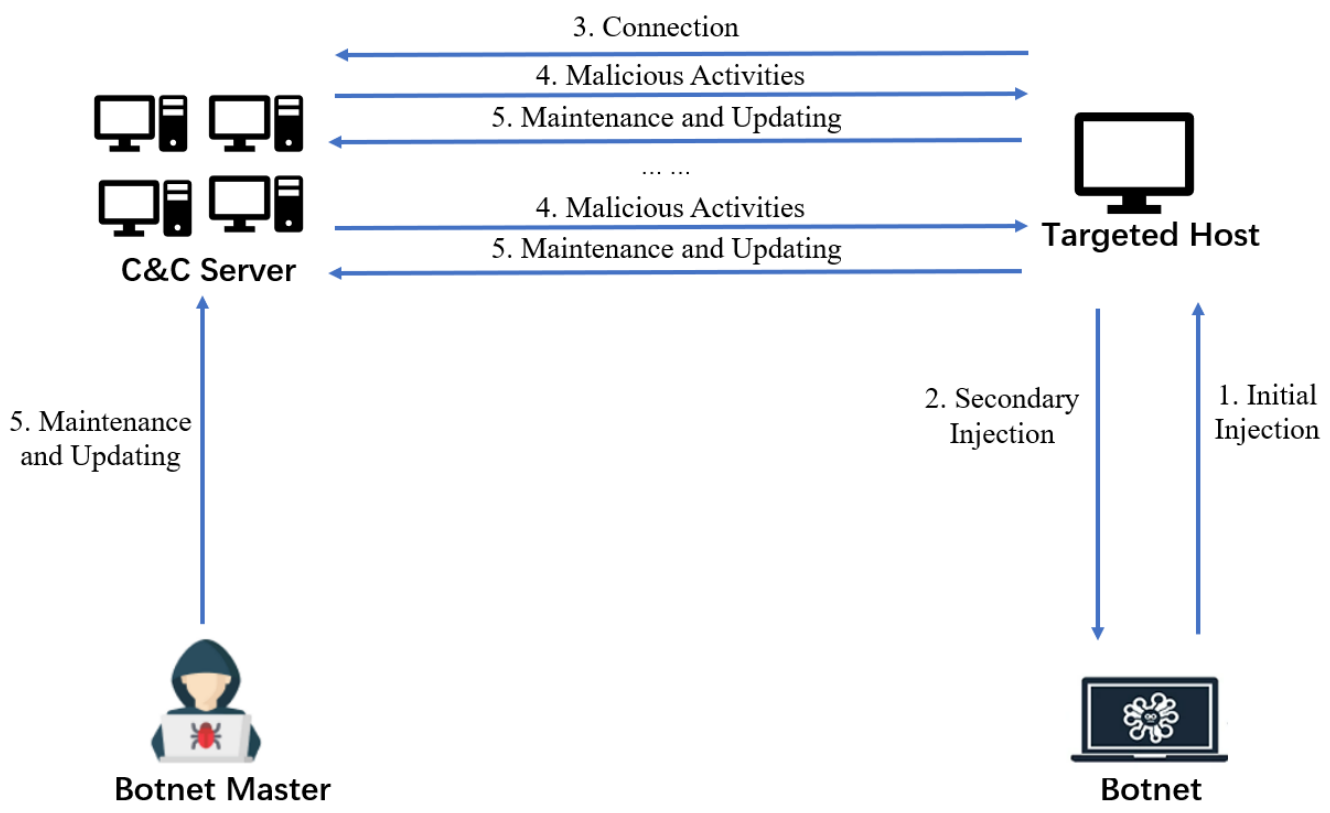

**FIGURE 9** The typical lifecycle of a botnet.

BotStop, a packet-based and ML-based botnet detection solution aimed at testing the incoming and outgoing network traffic in an IoT device to stop botnet infections, was introduced by Mohammed in [184]. The suggested method additionally emphasized feature selection to utilize only seven features to train an extremely accurate ML classifier. The trained classifier surpassed all methods from similar work with an accuracy of 0.9976, an F1-Score of 0.9968, and a testing duration of 0.2250 μs. Besides, very low FN and FP rates of 0.21 percent and 0.31 percent were attained using the suggested approach as well. SHAP explanation is used to explain the proposed model to make the classifier prediction process transparent.

### 4) FRAUD

According to [185], during the tightest periods of the lockdown during the Covid-19 epidemic, there were observed rises in personal account hacking and online financial fraud. In the UK, fraud costs businesses and individuals £130 billion per year, while it costs the worldwide economy $3.89 trillion [186]. Therefore, to deal with this issue, numerous financial services, have the potential to benefit from the use of AI systems to defend against fraud attacks. However, there are still practical challenges with the complete implementation of AI methods, and some focus on comprehending and being able to explain the judgments and predictions produced by complicated models by XAI [187].

Ismini *et al.* [187] investigated explanations for fraud detection by both supervised and unsupervised models using two of the most used techniques, LIME and SHAP. The open source IEEE-CIS Fraud Detection dataset [188] was tested on 8 popular supervised and unsupervised AI models including Naive Bayes, Logistic Regression, Decision Tree, Random Forest, Gradient Boosting, Neural Network, Autoencoder, and Isolation Forest whereas LIME and SHAP provided explanations for the detection results of each model respectively. It was noticed that while SHAP gives more reliable explanations, LIME is faster. Therefore, this paper suggested that combining the two approaches may be advantageous, with SHAP being used to facilitate regulatory compliance and LIME being used to offer real-time explanations for fraud prevention and model accuracy analysis.

David *et al.* [189] investigated how existing XAI algorithms may be used to explain specific predictions for prescriptive solutions and derive more information about the causes of cyber-fraud in the iGaming industry. ML algorithms including RF, LGB, DT, and LR were utilized to analyze a dataset with a sample size of 197,733. Besides, this study also proved the existence of data drift and suggested monthly retraining for the model to remain consistently updated. Furthermore, to identify the features that contributed most significantly to that particular case and to quantify that same contribution, this study employed locally faithful explanations. These explanations take the form of mathematical inequalities that reflect feature conditions, and each condition is assigned a relative strength. One of the research's limitations would be the manually labeled dataset, which could have added bias and human error to our analysis.

XFraud, an explainable fraud transaction prediction framework composed of a detector and an explainer, was presented by Susie *et al.* in [190]. A heterogeneous GNN model for transaction fraud detection was proposed and tested on industrial-scale datasets. Heterogeneity in transaction graphs was captured whereas the presented methodology outperformed previous models HGT [191] and GEM [192]. Besides, the weights learned by the GNNExplainer and the edge weights calculated using centrality measures were compared and traded off to compute a hybrid explainer in XFraud. The computed hybrid XFraud explainer calculated the contributions of its surrounding node types and edges and also paid attention to global topological aspects discovered by centrality metrics.

XAI methods can also be utilized to improve the performance of the fraud detection models. In [193], Khushnaseeb *et al.* proposed SHAP_Model based on the autoencoder for network fraud detection using SHAP values, implemented in a subset of the CICIDS2017 dataset and achieved overall accuracy and AUC of 94% and 96.9% respectively. The top 30 features with the highest SHAP values, playing a more significant role in causing abnormal behavior in fraud detection than any other features, were employed to build the SHAP_Model. Experimental results demonstrated that the SHAP_Model outperformed the model based on all features and the model based on 39features extracted by unsupervised learning.

Yongchun *et al.* [194] proposed a Hierarchical Explainable Network (HEN) to represent user behavior patterns, which could help with fraud detection while also making the inference process more understandable. Furthermore, a transfer framework was suggested for knowledge transfer from source domains with sufficient and mature data to the target domain to address the issue of cross-domain fraud detection.

A novel fraud detection algorithm called FraudMemory was proposed in [195] by Kunlin *et al.* This methodology used memory networks to enhance both performance and interpretability while using a novel sequential model to capture the sequential patterns of each transaction. Besides, memory components were incorporated in FraudMemory to possess high adaptability to the existence of the concept drift. The precision and AUC of the FraudMemory model were 0.968 and 0.969 respectively and performed better than any other methods for comparison including SVM, DNN, RF, and GRU.

Based on a real-world dataset and a simulated dataset, Zhiwen and Jianbin [196] proposed an explainable classification approach within the multiple instance learning (MIL) framework that deployed the AP clustering method in the self-training LSTM model to obtain a precise explanation. The experimental results indicated that the presented methodology surpassed the other 3 benchmark classifiers including AP, SVM, and RF in both 2 datasets. Only a few classification methods that can produce a straightforward casual explanation is the one used in this study.

Wei *et al.* [197] proposed a DL-based behavior representation framework for clustering to detect fraud in financial services, called FinDeepBehaviorCluster. Time attention-based Bi-LSTM was used to learn the embedding of behavior sequence data whereas handcrafted features were deployed to provide explanations. Then a GPU-optimized HDBSCAN algorithm called pHDBSCAN is used for clustering transactions with similar behaviors. The proposed pHDBSCAN has demonstrated comparable performance to the original HBDSCAN in experiments on two real-world transaction data sets but with hundreds of times greater computation efficiency.

#### 5) PHISHING

Phishing refers to fake email messages that look to be sent by a well-known company. The intention is to either download malicious software onto the victim's computer or steal sensitive data from it, including credit card numbers and login credentials. Phishing is a form of online fraud that is gaining popularity [198].

Yidong *et al.* [199] proposed a multi-modal hierarchical attention model (MMHAM) that, for phishing website detection, jointly learned the deep fraud cues from the three main modalities of website content including URLs, text, and image. Extracted features from different contents would be aligned representations in the attention layer. This methodology is self-explained because content distributed with the most attention would be regarded as the most important content contributing to the final decision.

Paulo *et al.* [200] utilized LIME and EBM explanation techniques based on malicious URLs for a phishing experiment on a publicly available dataset Ebbu2017 [201]. EBM, Random Forest, and SVM classifiers rated accuracy of 0.9646, 0.9732, and 0.9469 respectively on the tested database. The empirical evidence supported that the models could accurately categorize URLs as phishing or legitimate, and they also added explainability to these ML models, improving the final classification outcome.

Visual explanations of the phishing detection system attracted attention in the work of Yun *et al.* [202] as well. The proposed phishing website detection method Phishpedia solved the challenging issues of logo detection and brand recognition in phishing website detection. Both high accuracy and little runtime overhead are attained via Phishpedia. And most crucially, unlike conventional methods such as EMD, PhishZoo, and LogoSENSE, Phishpedia does not demand training on any specific phishing samples. Moreover, Phishpedia was implemented with the CertStream service, and in just 30 days, we found 1,704 new genuine phishing websites, far more than other solutions. In addition, 1,133 of these were not flagged by any engines in VirusTotal.

Rohit *et al.* [203] proposed an anti-phishing method that utilizes persuasion cues and investigated the effectiveness of persuasion cues. Three ML models were developed with pertinent gain persuasion cues, loss persuasion cues, and combined gain and loss persuasion cues, respectively, to respond to the research questions. We then compare the results with a baseline model that does not take the persuasion cues into account. The findings demonstrate that the three phishing detection models incorporating pertinent persuasion cues considerably outperform the baseline model in terms of F1-score by a range of 5% to 20%, making them effective tools for phishing email detection. In addition, the use of the theoretical perspective can aid in the creation of models that are comprehensible and can understand black-box models.

#### 6) NETWORK INTRUSION

An unauthorized infiltration into a computer in your company or an address in your designated domain is referred to as a network intrusion. On the other hand, Network Intrusion Detection Systems (NIDSs) are defined as monitoring network or local system activity for indications of unusual or malicious behavior that violates security or accepted practices [36]. Recently, many works have adopted ML and DL algorithms for building efficient NIDSs. In addition, cyber security experts also consider introducing explainability to the black-box AI systems to make the NISDs more robust and many have tried with XAI [204].

Pieter *et al.* [204] proposed a two-staged pipeline for robust network intrusion detection, which deployed XGBoost in the first phase and Autoencoder in the second phase. SHAP method was implemented to explain to the first stage model whereas the explanation results were utilized in the second stage to train the autoencoder. Experiments in the public corpus NSL-KDD [105] showed that the proposed pipeline can outperform many state-of-the-art efforts in terms of accuracy, recall, and precision with 93.28%, 97.81%, and 91.05% respectively on the NSL-KDD dataset while adding an extra layer of explainability.

ROULETTE, an explainable network intrusion detection system for neural attention multi-output classification of network traffic data was introduced by Giuseppina *et al.* in [205]. Experimentations were performed on two benchmark datasets, NSL-KDD [105] and UNSW-NB15 [113] to demonstrate the effectiveness of the proposed neural model with attention. The additional attention layer enables users to observe specific network traffic characteristics that are most useful for identifying particular intrusion categories. Two heatmaps depicting the ranked average feature relevance of the flow characteristics in the attention layer of the above 2 datasets were provided to show the explanation.

Zakaria *et al.* [206] designed a novel DL and XAI-based system for intrusion detection in IoT networks. Three different explanation methods including LIME, SHAP, and RuleFit were deployed to provide local and global explanations for the single output of the DNN model and the most significant features conducted to the intrusion detection decision respectively. Experiments were operated on NSL-KDD [105] and UNSW-NB15 [113] datasets and the performance results indicated the proposed framework's effectiveness in strengthening the IoT IDS's interpretability against well-known IoT assaults and assisting cybersecurity professionals in better comprehending IDS judgments.

Yiwen *et al.* [207] presented an intrusion detection system aimed at detecting malicious traffic intrusion in networks such as flood attacks and Ddos attacks. This method was XAI-based and deployed both neural networks and tree models. It is noticeable that this approach decreased the number of convolution layers in the neural work to enhance the model's explainability whereas the accuracy performance of the model was not sacrificed. XGBoost was implemented to process the prediction outputs of the neural network and the processed results would be fed to LIME and SHAP for further explanations.

A novel intrusion detection system known as BiLSTM-XAI was presented by S. Sivamohan *et al.* in [208]. Krill herd optimization (KHO) algorithm was implemented to generate the most significant features of two network intrusion datasets, NSL-KDD [105] and Honeypot [209], to reduce the complexities of BiLSTM model and thus enhance the detection accuracy and explainability. The obtained detection rate of Honeypot is 97.2% and the NSL-KDD dataset is 95.8% which was superior and LIME and SHAP were deployed to explain the detection decisions.

Hong *et al.* [210] suggested a network intrusion detection framework called FAIXID making use of XAI and data cleaning techniques to enhance the explainability and understandability of intrusion detection alerts. The proposed framework will help cyber analysts make better decisions because false positives will be quickly eliminated. Five functional modules were identified in FAIXID framework: the pre-modeling explainability model, the modeling module, the post-modeling explainability module, the attribution module, and the evaluation module. XAI algorithms including Exploratory Data Analysis (EDA), Boolean Rule Column Generation(BRCG), and Contrastive Explanations Method (CEM) were deployed in the pre-modeling explainability model, the modeling module, and the post-modeling explainability module respectively to provide cybersecurity analysts comprehensive and high-quality explanations about the detection decisions made by the framework. On the other hand, collecting analysts' feedback through the evaluation module to enhance the explanation models by data cleaning also proved effective in this work as well.

Shraddha *et al.* [211] proposed a system where the relations between features and system outcome, instance-wise explanations, and local and global explanations aid to get relevant features in decision making were identified to help users to comprehend the patterns that the model has learned by looking at the generated explanations. If the learned patterns are incorrect, they can alter the dataset or choose a different set of features to ensure that the model learns the correct patterns. XAI methods including SHAP, LIME, Contrastive Explanations Method (CEM), ProtoDash, and Boolean Decision Rules via Column Generation (BRCG) were implemented at different stages of the framework so that the neural network not being a black box. The experiment was performed on the dataset NSL-KDD [105] and the proposed framework was applied to generate explanations from different perspectives.

The Decision Tree algorithm was utilized by Basim *et al.* in [212] to enhance trust management and was compared with other ML algorithms such as SVM. By applying the Decision Tree model for the network intrusion of benchmark dataset NSL-KDD [105], three tasks were performed: ranking the features, decision tree rule extraction, and comparison with the state-of-the-art algorithms. The ranking of network features was listed and it is noticeable that not all features contributed to the decision of intrusion. Besides, the advantages of the Decision Tree algorithm compared with other popular classifiers, being computationally cheaper and easy to explain were also demonstrated in this work.

Syed *et al.* [213] suggested an Intrusion Detection System that used the global explanations created by the SHAP and Random Forest joint framework to detect all forms of malicious intrusion in network traffic. The suggested framework was composed of 2 stages of Random Forest classifiers and one SHAP stage. SHAP provided explanations for the outcome of the initial Random Forest classifier and one decision of the first Random Forest classifier with low credibility would be reassessed by the secondary classifier.

This three-stage based architecture can increase user trust while filtering out all cloaked dangerous network data by introducing transparency to the decision-making process. CSE-CIC IDS 2018 [214] dataset was utilized to evaluate the performance of the proposed framework and the presented architecture produced accuracy rates of 98.5 percent and 100 percent, respectively on the test dataset and adversarial samples.

Tahmina *et al.* [215] proposed an XAI-based ML system to detect malicious DoH traffic within DNS over HTTPS protocol. publicly available CIRA-CIC-DoHBrw-2020 dataset [216] was utilized in the testing of the proposed Balanced and Stacked Random Forest framework and other ML algorithms including Gradient Boosting and Generic Random Forest. The suggested approach in this work got slightly greater precision (99.91 percent), recall (99.92 percent), and F1 score (99.91 percent) over other methods for comparison. Additionally, feature contributions to the detection results were also highlighted with the help of the SHAP algorithm. The limitation of this framework would be the inconsideration of DGA-related DoH traffic from other HTTPS traffic.

#### 7) DOMAIN GENERATION ALGORITHMS (DGA)

DGAs are a type of virus that is frequently used to generate a huge number of domain names that can be utilized for evasive communication with Command and Control (C2) servers. It is challenging to prohibit harmful domains using common approaches like blacklisting or sink-holing due to the abundance of unique domain names. A DGA's dynamics widely used a seeded function. Deterring a DGA strategy presents a hurdle because an administrator would need to recognize the virus, the DGA, and the seed value to filter out earlier dangerous networks and subsequent servers in the sequence. The DGA makes it more challenging to stop unwanted communications because a skilled threat actor can sporadically switch the server or location from which the malware automatically calls back to the C2 [217]. Therefore, blacklisting and other conventional malware management techniques fall short in combating DGA attacks and many ML classifiers have been suggested. These classifiers allow for the identification of the DGA responsible for the creation of a given domain name and consequently start targeted remedial actions. However, it's challenging to assess the inner logic due to the black box aspect and the consequent lack of confidence makes it impossible to use such models.

Franziska *et al.* [218] proposed a visual analytics framework that offers clear interpretations of the models created by DL model creators for the classification of DGAs. The activations of the model's nodes were clustered, and decision trees were utilized to illuminate these clusters. The users can examine how the model sees the data at different layers in conjunction with a 2D projection. A drawback of the proposed strategy is that although the decision trees can provide a possible explanation for the clusters, this does not necessarily reflect how the model classifies this data, especially when there are numerous equally valid explanations.

EXPLAIN, a feature-based and contextless DGAs multiclass classification framework was introduced by Arthur *et al.* in [219] and compared with several state-of-the-art classifiers such as RNN, CNN, SVM, RF, and ResNet based on real-world datasets including DGArchive [220] and University Network [221]. After the ResNet-based techniques, the best model, EXPLAIN-OvRUnion, used 76 features and achieves the best F1-score. Moreover, Only 28 features were used by EXPLAIN-OvRRFE-PI and EXPLAIN-RFRFE-PI, which outperformed all feature-based strategies put out in previous work by a significant margin. Additionally, they outperformed the DL-based algorithms M-Endgame, M-Endgame.MI, and M-NYU in terms of F1-scores as well.

To address the issues of DGAs classification including which traffic should be trained in which network and when, and how to measure resilience against adversarial assaults, Arthur *et al.* [222] proposed two ResNets-based DGAs detection classifiers, one for binary classification and the other for multiclass classification. Experiments on real-world datasets demonstrated that the proposed classifier performed at least comparably to the best state-of-the-art algorithms for the binary classification test with a very low false positive rate, and significantly outperformed the competition in the extraction of complex features. In addition, for the multiclass classification problem, the ResNet-based classifier performed better than previous work in attributing AGDs to DGAs for the multiclass classification problem, achieving an improvement of nearly 5 percent in F1-score while requiring 30 percent less training time than the next best classifier. In the explainability analysis, it was also highlighted that some of the self-learned properties employed by the DL-based systems.

#### 8) DENIAL-OF-SERVICE (DOS)

The Internet is seriously threatened by denial-of-service (DoS) assaults, and numerous protection measures have been suggested to address the issue. DoS attacks are ongoing attacks in which malicious nodes produce bogus messages to obstruct network traffic or drain the resources of other nodes [223]. As the DoS attacks become increasingly complicated in the past years, conventional Intrusion Detection Systems (IDS) are finding it increasingly challenging to identify these newer, more sophisticated DoS attacks because they use more complicated patterns. To identify malicious DoS assaults, numerous ML and DL models have been deployed. Additionally, for the goal of model transparency, XAI methods that investigate how features contribute to or impact an algorithm-based choice can be helpful [224].

**TABLE 6.** Details of XAI applications in defending mechanisms against different categories of cyber attacks.

| Cyber attack types | Reference | Learning models | Year | XAI techniques | | | | | | | | | | |
|---|---|---|---|---|---|---|---|---|---|---|---|---|---|---|
| | | | | **Local** | **Global** | **Model-specific** | **Model-agnostic** | **Post-hoc** | **Intrinsic** | **Text** | **Visual** | **Arguments** | **Models** | **XAI methods** |
| Malware | [150] | SVM and RF | 2018 | √ | √ | | √ | | √ | | | √ | | gradient |
| | [154] | DNN | 2020 | √ | | √ | | | √ | | √ | √ | | heatmap |
| | [157] | CNN | 2020 | √ | √ | √ | | √ | | | √ | √ | | Grad-CAM |
| | [158] | DNN | 2021 | √ | | √ | | √ | | | | | √ | Generated trees |
| | [159] | CNN | 2021 | √ | | | | √ | | √ | √ | | | Grad-CAM , heatmap |
| | [160] | DT | 2016 | | √ | √ | | √ | | | | √ | | Self explainable |
| | [161] | RF, LR, DT, GNB, and SVM | 2022 | √ | | | √ | √ | | | √ | √ | | SHAP |
| | [162] | CNN | 2021 | √ | | | √ | √ | | | √ | √ | | LIME |
| Spam | [168] | XGBoost | 2019 | √ | | | √ | √ | | | √ | √ | | SHAP |
| | [169] | NB and RF | 2020 | | | √ | | | | | | √ | | Self explainable |
| | [171] | RNN and CNN | 2021 | √ | | | √ | √ | | | √ | √ | | LIME |
| Botnet | [178] | RF, NB, and LR | 2022 | √ | | | √ | √ | | | √ | √ | | LIME and SHAP |
| | [179] | DBSCAN | 2019 | | √ | √ | | | √ | | | √ | √ | Self explainable |
| | [181] | VAEs and LSTM | 2019 | | | √ | | √ | | | √ | | | Visualized tools |
| | [182] | DT | 2018 | | | √ | | | √ | | | | √ | Self explainable |
| | [183] | DCNN | 2022 | √ | | | √ | √ | | | √ | √ | | SHAP |
| | [184] | ML | 2022 | √ | | | √ | √ | | | √ | √ | | SHAP |
| Fraud | [187] | Autoencoder, NB, RF and DT | 2021 | √ | | | √ | √ | | | √ | √ | | LIME and SHAP |
| | [189] | RF, LGB, DT, and LR | 2021 | √ | | | √ | √ | | | √ | | | Local features |
| | [190] | GNN | 2022 | √ | √ | | √ | √ | | | | √ | | GNN Explainer |
| | [193] | Autoencoder | 2021 | √ | | | √ | √ | | | √ | √ | | Kernel SHAP |
| | [194] | Transfer Learning | 2020 | | | √ | | | √ | | √ | √ | | HEN |
| | [195] | Sequential modeling | 2019 | | | √ | | | √ | | | | √ | Fraud Memory |
| | [196] | AP Clustering and LSTM | 2021 | | | √ | | | √ | | | | √ | MIL |
| | [197] | Bi-LSTM and pHDBSCAN | 2021 | | | √ | | | √ | | √ | | √ | Feature extraction |
| Phishing | [199] | MMHAM | 2022 | | √ | √ | | | √ | √ | √ | √ | | Self explainable |
| | [200] | RF and SVM | 2021 | √ | | | √ | √ | | | √ | √ | | LIME and EBM |
| | [202] | Phishpedia | 2021 | | | √ | | √ | | | √ | | | Visual explanation |
| | [203] | NB, LR, RF, and SVM | 2021 | | | √ | | √ | | | √ | | | Theoretical Perspective |
| Network Intrusion | [204] | XGBoost and autoencoder | 2022 | √ | | | √ | √ | | | √ | √ | | SHAP |
| | [205] | Neural network and attention | 2022 | √ | | √ | | | √ | | √ | | | Self explainable |
| | [206] | DNN | 2022 | √ | √ | | √ | √ | | | √ | √ | | LIME, SHAPE, and RuleFit |
| | [207] | CNN, LSTM, and XGBoost | 2022 | √ | | | √ | √ | | | √ | √ | | LIME and SHAP |
| | [208] | BiLSTM | 2022 | √ | √ | | √ | √ | | | √ | √ | | KHO, LIME, and SHAP |
| | [210] | DNN | 2021 | √ | √ | | √ | √ | √ | | √ | √ | | EDA, BRCG, and CEM |
| | [211] | DNN | 2021 | √ | √ | | √ | √ | √ | | √ | √ | | SHAP, LIME, and BRCG |
| | [212] | DT | 2021 | √ | | √ | | | √ | | | √ | √ | Self explainable |
| | [213] | RF | 2021 | √ | | | √ | √ | | | √ | √ | | SHAP |
| | [215] | Stacked RF | 2022 | √ | | | √ | √ | | | √ | √ | | SHAP |
| Domain | [219] | CNN and RNN | 2020 | | √ | | √ | | √ | | √ | √ | | Clustering and DT |
| | [220] | RNN, CNN, | 2021 | | √ | | √ | √ | | | | √ | | EXPLAIN |
| Generation Algorithms (DGA) | | SVM, RF, and ResNet | | | | | | | | | | | | |
| | [222] | ResNet | 2020 | | √ | | √ | √ | | | | √ | | Self explainable |
| Denial-of-Service (DoS) | [225] | XGBoost | 2022 | √ | | | √ | √ | | | √ | √ | | SHAP |
| | [226] | ML | 2021 | | √ | | √ | √ | | | | √ | | TCAV |
| | [228] | DNN | 2018 | | √ | √ | | √ | | √ | √ | √ | | DNN Explanation Generator |

Boryau *et al.* [225] introduced CSTITool, a CICFlowMeter-based flow extraction to feature extraction to enhance the performance of the ML DoS attack detection model. CICFlowMeter translated the flow data from packets for the model's training. The size of the data was significantly reduced during this process, which decreased the need for data storage. Hacker attack data including Network Service Scanning, Endpoint DoS, Brute Force, and Remote Access Software from the dataset CIC-IDS2017 network flow data of malware from the dataset CSTI-10 were utilized to train the XGBoost model. The outcome demonstrated that the performance measurements can be enhanced by using the additional descriptive flow statistics produced by CSTITool. For instance, Rig's Precision and Recall increased by 1.23% and 1.59% respectively. Moreover, XAI method SHAP was deployed to further explore the relationship between cyberattacks and network flow variables to better understand how the model produced predictions.

In the context of DoS attack, Rendhir *et al.* [226] analyzed the strategic decisions based on the KDD99 dataset [227] with the XAI method of Testing with ConceptActivation Vectors (TCAV). The approach investigates the connection between the strategic choice, autonomous agent's objective, and dataset properties. TCAVQ scores are obtained from the KDD99 dataset for various DoS attacks and regular traffic. The relationship between the goal availability and the strategies TerminateConnection and AllocateMoreResources is determined using the TCAVQ scores. In the event of cyberattacks, the analysis is performed to support the choice of the plan or, if necessary, a change in the strategy.

Kasun *et al.* [228] described the framework for explainable DNNs-based DoS anomaly detection in process monitoring. The user was given post-hoc explanations for DNN predictions in the framework that is currently being used. Based on the DoS attack benchmark dataset NSL-KDD [105], experiments were implemented on several DNN architectures, and it was found that on the test dataset, DNNs were able to yield accuracies of 97%. Besides, according to experimental findings, while classified as DoS, DNNs could also provide a higher relevance to the number of connections, connection frequency, and volume of data exchanged. Therefore, this framework improves human operators' confidence in the system by reducing the opaqueness of the DNN-based anomaly detector.

## B. XAI FOR CYBER SECURITY IN INDUSTRIAL APPLICATIONS

In this subsection, we aim to present a comprehensive overview of XAI studies for the cyber security of different industrial areas, as shown in Figure 9. And the details of these XAI implementations for cyber security in distinct industries are shown in Table 7 as well.

### 1) XAI FOR CYBER SECURITY OF HEALTHCARE

The use of big data, cloud computing, and IoT creates a modern, intelligent healthcare industry. The use of the Internet of Things, cutting-edge manufacturing technologies, software, hardware, robots, sensors, and other sophisticated information technologies, improves data connectivity. Information and communication technology advancements enhance the quality of healthcare by transforming conventional healthcare organizations into smart healthcare [229]. With the increasingly significant role of AI in healthcare, there are growing concerns about the vulnerabilities of the smart healthcare system. Smart healthcare is a prime target for cybercrime for two main reasons: a vast supply of valuable data and its defenses are porous. Health information theft, ransomware attacks on hospitals, and potential attacks on implanted medical equipment are all examples of cyber security breaches. Breaches can undermine smart healthcare systems, erode patient trust, and endanger human life [230].

XAI comes into the picture as the smart healthcare system demands transparency and explainability to decrease the increasing vulnerabilities of the smart healthcare system due to the increasingly connected mobile devices, more concern for patients' monitoring, and more mobile consumer devices. There are many studies currently on implementing the XAI framework to address the issue of privacy and security of the smart healthcare system.

Devam *et al.* [231] introduced a study based on the heart disease dataset and illustrated why explainability techniques should be chosen when utilizing DL systems in the medical field. This study then suggested and described various example-based strategies, such as Anchors, Counterfactuals, Integrated Gradients, Contrastive Explanation Method, and Kernel Shapley, which are crucial for disclosing the nature of the model's black box and ensuring model accountability. These XAI approaches were compared with two benchmark XAI methods, LIME and SHAP, as well. It was concluded that these discussed XAI approaches all explained how different features contribute to the outputs of the model. They are intuitive, which helps in the process of understanding what the black box model thinks and explains the model's behavior.

BrainGNN, an explainable graph neural network (GNN) based framework to analyze functional magnetic resonance images (fMRI) and identify neurological biomarkers was proposed by Xiaoxiao *et al.* [232]. Motivated by the requirements for transparency and explainability in medical image analysis, the proposed BrainGNN framework included ROI-selection pooling layers (R-pool) that highlight prominent ROIs (nodes in the graph) so that which ROIs are crucial for prediction could be determined. By doing so, the advantage of the BrainGNN framework could be the allowance of users to interpret significant brain regions in multiple ways.

The chain of reasoning behind Computer Aided Diagnostics (CAD) is attracting attention to build trust in CAD decisions from complicated data sources such as electronic health records, magnetic resonance imaging scans, cardiotocography, etc. To address this issue, Julian *et al.* [233] presented a new algorithm, Adaptive-Weighted High Importance Path Snippets (Ada-WHIPS) to explain AdaBoost classification with logical and simple rules in the context of CAD-related data sets. The weights in the individual decision nodes of the internal decision trees of the AdaBoost model are redistributed especially by Ada-WHIPS. A single rule that dominated the model's choice is then discovered using a straightforward heuristic search of the weighted nodes. Moreover, according to experiments on nine CAD-related data sets, Ada-WHIPS explanations typically generalize better (mean coverage 15 percent to 68 percent) than the state of the art while being competitive for specificity.

A novel human-in-the-loop XAI system, XAI-Content based Image Retrieval (CBIR), was introduced by Deepak *et al.* in [234] to retrieve video frames from minimally invasive surgery (MIS) videos that are comparable to a query image based on content. MIS video frames were processed using a self-supervised DL algorithm to extract semantic features. The search results were then iteratively refined using an iterative query refinement technique, which utilized a binary classifier that has been trained online using user feedback on relevance. The saliency map, which provided a visual description of why the system deems a retrieved image to be similar to the query image, was produced using an XAI technique. The proposed XAI-CBIR system was tested using the publicly available Cholec80 dataset, which contains 80 films of minimally invasive cholecystectomy procedures.

#### 2) XAI FOR CYBER SECURITY OF SMART CITIES

As increasingly data-driven artificial intelligence services such as IoT, blockchain, and DL are incorporated into contemporary smart cities, smart cities are able to offer intelligent services for energy, transportation, healthcare, and entertainment to both city locals and visitors by real-time environmental monitoring [235]. However, smart city applications not only gather a variety of information from people and their social circles that are sensitive to privacy, but also control municipal services and have an impact on people's life, cyber security, cyber crime, and privacy problems about smart cities arise. To address this issue, XAI integration into IoT and AI-enabled smart city applications can help to address black-box model difficulties and offer transparency and explainability components for making useful data-driven decisions for smart city applications. Smart city applications are usually utilized in high-risk and privacy-sensitive scenarios. Therefore, it is crucial to establish an effective XAI approach to give authorities additional information about the justification, implications, potential throughput, and an in-depth explanation of background procedures to aid in final decision-making [236].

Roland *et al.* [237] introduced a tree-based method Gradient Boosted Regression Trees (GBRT) model in conjunction with the SHAP-value framework to identify and analyze major patterns of meteorological determinants of PM1 species and overall PM1 concentrations. SIRTA [238], a ground-based atmospheric observatory dataset for cloud and aerosol was utilized to experiment and the location for establishing this dataset was in the city of Paris. The findings of this study show that shallow MLHs, cold temperatures, and low wind speeds play distinct roles during peak PM1 events in winter. Under high-pressure synoptic circulation, northeastern wind input frequently intensifies these conditions.

One of the most demanded bus lines of Madrid was analyzed by Leticia *et al.* in [239] to make the smart city transport network more efficient by predicting bus passenger demand. The proposed method created an interpretable model from the Long Short Term Memory (LSTM) neural network that enhances the generated XAI model's linguistic interpretability without sacrificing precision using a surrogate model and the 2-tuple fuzzy linguistic model. The public transportation business can save money and energy by using passenger demand forecasting to plan its resources most effectively. This methodology can also be used in the future to forecast passenger demand for other forms of transportation (air, railway, marine).

Georgios *et al.* [240] proposed explainable models for early prediction of certification in Massive Open Online Courses (MOOCs) for Smart City Professionals. MOOCs have grown significantly over the past few years due to Covid-19 and tend to become the most common type of online and remote higher education. Several ML classification techniques such as Adaptive Boosting, Gradient Boosting, Extremely Randomized Trees, Random Forest, and Logistic Regression were utilized to build corresponding predictive models using PyCaret. And the XAI method SHAP summary plot was employed to the classifiers including LightGBM, GB, and RF. Furthermore, new classification models based only on the two most important features in each step gained from the SHAP summary plot. And the experimental results showed that the effectiveness of all methods was slightly improved for all metrics.

#### 3) XAI FOR CYBER SECURITY OF SMART FARMING

Smart farming refers to the use of cutting-edge technology in agriculture, including IoT, robots, drones, sensors, and geolocation systems. Big data, cloud computing, AI, and augmented reality are the engines of smart farming as well. However, the addition of several communication modules and AI models leaves the system open to cyber-security risks and threats to the infrastructure for smart farming [241]. And cyber attacks can harm nations' economies that heavily rely on agriculture. However, due to the black box nature of most AI models, users cannot understand the connections between features. This is crucial when the system is designed to simulate physical farming events with socioeconomic effects like evaporation [242]. Therefore, many researchers are working on the implementation potentials of XAI applied in smart farming cyber security.

Nidhi *et al.* [242] presented an IoT and XAI-based framework to detect plant diseases such as rust and blast in pearl millet. Parametric data from the pearl millet farmland at ICAR, Mysore, India was utilized to train the proposed Custom-Net DL Models, reaching a classification accuracy of 98.78% which is similar to state-of-the-art models including Inception ResNet-V2, Inception-V3, ResNet-50, VGG-16, and VGG-19 and superior to them in terms of reducing the training time by 86.67%. Additionally, the Grad-CAM is used to display the features that the Custom-Net extracted to make the framework more transparent and explainable.

To thoroughly assess the variables that can potentially explain why agricultural land is used for plantations of wheat, maize, and olive trees, Viana *et al.* [243] implemented an ML and agnostic-model approach to show global and local explanations of the most important variables. ML model Random Forest and XAI approach LIME were deployed for analysis and approximately 140 variables related to agricultural socioeconomic, biophysical, and bioclimatic factors were gathered. By applying the proposed framework, it is found that the three crop plantations in the research area's usage of agricultural land were explained by five major factors: drainage density, slope, soil type, and the ombrothermic index anomaly (for humid and dry years).

#### 4) XAI FOR CYBER SECURITY OF SMART FINANCIAL SYSTEM

The financial system has been rapidly altered by AI models, which offer cost savings and improved operational efficiency in fields like asset management, investment advice, risk forecasting, lending, and customer service [244]. On one hand, the ease of using AI in these smart financial systems provides efficiency for all parties involved, but on the other hand, the risk of cyberattacks on them is growing exponentially. Attackers have traditionally been motivated primarily by money, making smart financial systems their top choice of target. To combat the finance crime targeting smart financial systems, one of the primary priorities in the smart financial domain should be the implementation of XAI [245]. The reason behind this issue is that it is essential in these extremely sensitive areas such as Money Laundering detection and Corporate Mergers and Acquisitions to not only have a highly accurate and robust model but also to be able to produce helpful justifications to win a user's faith in the automated system.

Swati *et al.* [246] proposed a belief-rule-based automated AI decision-support system for loan underwriting (BRB). This system can take into account human knowledge and can employ supervised learning to gain knowledge from prior data. Factual and heuristic rules can both be accommodated by BRB's hierarchical structure. The significance of rules triggered by a data point representing a loan application and the contribution of attributes in activated rules can both be used to illustrate the decision-making process in this system. The textual supplied to rejected applicants as justification for declining requesters' loan applications might have been started by the progression of events from the factual-rule-base to the heuristic-rule-base.

A novel methodology for producing plausible counterfactual explanations for the Corporate Mergers and Acquisitions (M&A) Deep Transformers system was presented by Linyi *et al.* [247]. The proposed transformer-based classifier made use of the regularization advantages of adversarial training to increase model resilience. More significantly, a masked language model for financial text categorization that improved upon prior methods to measure the significance of words and guarantee the creation of credible counterfactual explanations was developed. When compared to state-of-art methods including SVM, CNN, BiGRU, and HAN, the results show greater accuracy and explanatory performance.

An interactive, evidence-based method to help customers understand and believe the output produced by AI-enabled algorithms was generated for analyzing customer transactions in the smart banking area by Ambreen [248]. A digital dashboard was created to make it easier to engage with algorithm results and talk about how the suggested XAI method can greatly boost data scientists' confidence in their ability to comprehend the output of AI-enabled algorithms. In the proposed model, a Probabilistic Neural Network (PNN) was utilized to classify the multi-class scenario of bank transaction classification.

#### 5) XAI FOR CYBER SECURITY OF HUMAN-COMPUTER INTERACTION (HCI)

HCI enables people to comprehend and engage with technology by establishing an effective channel of communication. And HCI's primary goal is to create interactions that take users' wants and abilities into account [249]. In the field of HCI, security and privacy have long been significant research concerns, where Usable Security has arisen as an interdisciplinary research area. On the other hand, HCI and AI emerge together in such a way that AI imitates human behavior to create intelligent systems, whereas HCI tries to comprehend human behavior to modify the machine to increase user experience, safety, and efficiency. However, from an HCI standpoint, there is no assurance that an AI system's intended users will be able to comprehend it. And according to the user-centered design (UCD), a design must offer an understandable AI that cyber-attacks the requirements and skills of the intended users (e.g., knowledge level). Therefore, the final objective of XAI in HCI should be to guarantee that target users can comprehend the outcomes, assisting them in becoming more efficient decision-makers [250].

Gaur *et al.* [251] utilized XAI methods including LIME and SHAP in conjunction with ML algorithms including Logistic Regression(80.87%), Support Vector Machine(85.8%), K-nearest Neighbour(87.24%), Multilayer Perceptron(91.94%), and Decision Tree(100%) to build a robust explainable HCI model for examining the mini-mental state for Alzheimer's disease. It is worth mentioning that the most significant features contributing to the Alzheimer's disease examing were different for the LIME-based framework and the SHAP-based framework. In contrast to nWBV's dominance of the LIME features, MMSE makes a significant contribution to Shapely values.

To fill the gap few publications on artistic image recommendation systems give an understanding of how users perceive various features of the system, including domain expertise, relevance, explainability, and trust, Vicente *et al.* [252] examed several aspects of the user experience with a recommender system of artistic photos from algorithmic and HCI perspectives. Three different recommender interfaces and two different Visual Content-based Recommender (VCBR) algorithms were employed in this research.

Q. Vera *et al.* [253] presented a high-level introduction of the XAI algorithm's technical environment, followed by a selective examination of current HCI works that use human-centered design, evaluation, and provision of conceptual and methodological tools for XAI. Human-centered XAI was highlighted in this research, and the emerged research communities of human-centered XAI were introduced in the context of HCI.

#### 6) XAI FOR CYBER SECURITY OF SMART TRANSPORTATION

The emergence of cutting-edge technologies including software-defined networks (SDNs), IIoT, Blockchain, AI, and vehicular ad hoc networks (VANETs) has increased operational complexity while smoothly integrating smart transportation systems [254]. However, it can experience security problems that leave the transportation systems open to intrusion. In addition, security concerns in transportation technology affect the AI model [255]. Major transportation infrastructures such as Wireless Sensor Networks (WSN), Vehicle-to-everything communication (V2X), VMS, and Traffic Signal Controllers (TSC) have either already been targeted or are still susceptible to hacking. To defend against these cyber attacks and prevent the potential cyber threats on the smart transportation system, AI-enabled intrusion detection systems are introduced recently. Although In the past few years, AI has made significant progress in providing effective performance in smart transportation systems, the XAI methods are still required as XAI could make it possible for the smart transportation system to monitor transportation details such as drivers' behaviour, accicent causes, and vechicles' conditions.

A ML approach to detect misbehaving vehicles in the Vehicular Adhoc Networks (VANET) was proposed by Harsh *et al.* [256]. In the smart VANET, the performance of each vehicle depends upon the information from other autonomous vehicles (AVs). Therefore, the misinformation from misbehaving vehicles would damage the entire VANET as a whole and detecting misbehaving would be significant to build a stable and safe VANET system. Vehicular reference misbehavior (VeReMi) dataset [257] was utilized in an ensemble learning using Random Forest algorithm and a decision tree-based algorithm and accuracy and F1 score of 98.43% and 98.5% were achieved respectively.

Shideh *et al.* [258] described a transportation energy model (TEM) that forecasts home transportation energy use using XAI technique LIME. Data from Household Travel Survey (HTS), which is utilized to train the artificial neural network accurately, has been deployed in TEM and high validation accuracy (83.4%) was developed. For certain traffic analysis zones (TAZs), the significance and impact (local explanation) of HTS inputs (such as household travel, demographics, and neighborhood data) on transportation energy consumption are studied. The explainability of the proposed TEM framework can help the home transportation energy distribution in two ways, including describing the local inference mechanisms on individual (household) predictions and assessing the model's level of confidence can be done using a broad grasp of the model.

C. Bustos *et al.* [259] provided an automated scheme for reducing traffic-related fatalities by utilizing a variety of Computer Vision techniques (classification, segmentation, and interpretability techniques). An explainability analysis based on image segmentation and class activation mapping on the same images, as well as an adaptation and training of a Residual Convolutional Neural Network to establish a danger index for each specific urban scene, are all steps in this process. This computational approach results in a fine-grained map of risk levels across a city as well as a heuristic for identifying potential measures to increase both pedestrian and automobile safety.

### C. CYBER THREATS TARGETING XAI AND DEFENSIVE APPROACHES

In the above sections, the applications of XAI in different areas to defend against different cyber threats have been discussed. Nevertheless, although XAI could be effective in protecting other areas and models by providing transparency and explainability, XAI models themselves would face cyber threats as well. Both the AI models deployed and the explainability part could be vulnerable to cyber attacks.

Some cyber attackers even utilize the explainable characteristics to attack the XAI model. Therefore, we deem it necessary to review the cyber threats targeting XAI and corresponding defensive approaches against them in this review.

Apart from the different parts that conventional AI models need to protect, including samples, learning models, and the interoperation processes, the explainable part of XAI-based models should be paid attention to as well. The following researches describe some cyber attacks targeting XAI models using different approaches from different perspectives.

A novel black box attack was developed by Aditya *et al.* [260] to examine the consistency, accuracy, and confidence security characteristics of gradient-based XAI algorithms. The proposed black box attack focused on two categories of attack: CI and I attack. While I attack attempts to attack the single explainer without affecting the classifier's prediction given a natural sample, the CI attack attempts to simultaneously compromise the integrity of the underlying classifier and explainer. It is demonstrated that the effectiveness of the attack on various gradient-based explainers as well as three security-relevant data sets and models through empirical and qualitative evaluation.

Thi-Thu-Huong *et al.* [261] proposed a robust adversarial image patch (AIP) that alters the causes of interpretation model prediction outcomes and leads to incorrect deep neural networks (DNNs) model predictions, such as gradient-weighted class activation mapping. Four tests pertaining to the suggested methodology were carried out on the ILSVRC image dataset. There are two different kinds of pre-trained models (i.e., feature and no feature layer). The Visual Geometry Group 19-Batch Normalization (VGG19-BN) and Wide Residual Networks models, in particular, were used to test the suggested strategy (Wide ResNet 101). Two more pre-trained models: Visual Geometry Group 19 (VGG19) and Residual Network (ResNext 101 328d), were also deployed whereas masks and heatmaps from Grad-CAM results were utilized to evaluate the results.

Tamp-X, a unique approach that manipulates the activations of powerful NLP classifiers was suggested by Hassan *et al.* [262], causing cutting-edge white-box and black-box XAI techniques to produce distorted explanations. Two steps were carried out to evaluate state-of-art XAI methods, including the white-box InteGrad andSmoothGrad, and the black-box—LIME and SHAP. The first step was to randomly mask keywords and observe their impact on NLP classifiers whereas the second step was to tamper with the activation functions of the classifiers and evaluate the outputs. Additionally, three cutting-edge adversarial assaults were utilized to test the tampered NLP classifiers and it was found that the adversarial attackers have a much tougher time fooling the tampered classifiers.

Slack *et al.* [263] provided a unique scaffolding method that, by letting an antagonistic party create any explanation they want, effectively masks the biases of any given classifier. Extensive experimental testing using real data from the criminal justice and credit scoring fields showed that the proposed fooling method was successful in producing adversarial classifiers that can trick post-hoc explanation procedures, including LIME and SHAP, with LIME being found to be more susceptible than SHAP. In detail, it was demonstrated how highly biased (racist) classifiers created by the proposed fooling framework can easily deceive well-liked explanation techniques like LIME and SHAP into producing innocent explanations which do not reflect the underlying biases using extensive evaluation with numerous real-world datasets (including COMPAS [264]).

Simple, model-agnostic, and intrinsic Gradient-based NLP explainable approaches are considered faithful compared with other state-of-art XAI approaches including SHAP and LIME. However, Junlin *et al.* [265] show how the gradients-based explanation methods can be fooled by creating a FACADE classifier that could be combined with any particular model having deceptive gradients. Although the gradients in the final model are dominated by the customized FACADE model, the predictions are comparable to those of the original model. They also demonstrated that the proposed method can manipulate a variety of gradient-based analysis methods: saliency maps, input reduction, and adversarial perturbations all misclassify tokens as being very significant and of low importance.

On the other hand, to defend against these cyber threats targeting XAI models, researchers also developed several defensive approaches, divided into three main categories: modifying the training process and input data, modifying the model network, and sing auxiliary tools.

Gintare *et al.* [266] assessed how JPG compression affects the categorization of adversarial images. Experimental tests demonstrated that JPG compression could undo minor adversarial perturbations brought forth by the Fast-Gradient-Sign technique. JPG compression could not undo the adversarial perturbation, nevertheless, if the perturbations are more significant. In this situation, neural network classifiers' strong inductive bias cause inaccurate yet confident misclassifications.

Ji *et al.* [267] present DeepCloak, a defense technique. DeepCloak reduces the capacity an attacker may use to generate adversarial samples by finding and eliminating pointless characteristics from a DNN model, increasing the robustness against such adversarial attacks. In this work, the mask layer, inserted before processing the DNN model, encoded the discrepancies between the original images and related adversarial samples, as well as between these images and the output features of the preceding network model layer.

Pouya *et al.* [268] Defense-GAN, a novel defense technique leveraging GANs to strengthen the resilience of classification models against adversarial black-box and white-box attacks. The proposed approach was demonstrated to be successful against the majority of frequently thought-of attack tactics without assuming a specific assault model. On two benchmark computer vision datasets, we empirically demonstrate that Defense-GAN consistently offers acceptable defense while other approaches consistently struggled against at least one sort of assault.

## VI. ANALYSIS AND DISCUSSION

### A. CHALLENGES OF USING XAI FOR CYBER SECURITY

We have reviewed the state-of-art XAI techniques utilized in the defense of different cyber attacks and the protection of distinct industrial cyber security domains. It is noticeable that although XAI could be a powerful tool in the application of different cyber security domains, XAI faces certain challenges in its application of cyber security. And in this section, we will discuss these challenges.

#### 1) DATASETS

An overview of the famous and commonly used datasets of different cyber attacks and distinct industries was provided in Table 4 and Table 5 respectively. However, there is a severe issue with the most used cyber security datasets, i.e. many datasets are not updated in certain directions. This phenomenon may be caused by privacy and ethical issues. Therefore, the most recent categories of cyber attacks were not included in the public cyber attack datasets, which would lead to inefficiency in the training of the XAI applications in the establishment of cyber attack defensive mechanisms. Although the industrial datasets in areas such as healthcare, smart agriculture, and smart transportation include more recent samples than the datasets for cyber attacks, these datasets should be updated as well because cyber attacks are becoming more sophisticated and diverse these days. Another issue with the currently available datasets is that these datasets usually lack a large volume of data available for the training of XAI methods, which will decrease both the performance and the explainability of the XAI approaches. Another reason behind this situation is that some of the information related to cyber attacks and cyber industries is redundant and unbalanced. Other than that, the heterogeneity of samples collected in these datasets is a challenge for the XAI models as well. The number of features and categories varies for each dataset and some datasets are composed of human-generated cyber attacks rather than exhibiting real-world and latest attacks. These problems highlight the challenge that the most recent benchmark datasets with a massive amount of data for training and testing and a balanced and equal number of attack categories are still to be identified.

#### 2) EVALUATION

Evaluation measure for XAI systems is another important factor in the application of XAI approaches for cyber security. When evaluating the performance of the established XAI-based cyber security systems, several conventional evaluation metrics including F1-Score, Precision, and ROC could be utilized to measure the performance of the proposed mechanisms. However, when applying XAI methods in the cyber security domains, measurements to evaluate the accuracy and completeness of explanations from the XAI systems are required. In general, the evaluation measurements of XAI systems should be able to assess the quality, value, and satisfaction of explanations, the enhancement of the users' mental model brought about by model explanations, and the impact of explanations on the effectiveness of the model as well as on the users' confidence and reliance. Unfortunately, the findings derived from the above reviews of this survey demonstrate the challenge that: more generic, quantifiable XAI system evaluation measurements are required to support the community's suggested XAI explainability measuring techniques and tools. Popular XAI explanation evaluation measurements can be divided into two main categories: user satisfaction and computational measurements. However, user satisfaction-based evaluation approaches are dependent on user feedback or interview, which may cause privacy issues for many cyber security problems. On the other hand, for computational measurements, many researchers utilize inherently interpretable models [56] (e.g., linear regression and decision trees) to compare with the generated explanations. Nevertheless, there are no benchmark comparison models for this evaluation approach, and the users' understanding of the explanation could not be reflected. Besides, the XAI evaluation systems lack measurements focusing on some other significant factors of the cyber security domain including computational resources as well as computational power. In conclusion, it is necessary to take into account a set of agreed-upon standard explainability evaluation metrics for comparison to make future improvements for XAI applications in cyber security.

#### 3) CYBER THREATS FACED BY XAI MODELS

As we discussed in Section V, although XAI methods can provide transparency and explainability to AI-enabled systems to prevent cyber threats, the current XAI models are facing many cyber attacks targeting the vulnerabilities of the explanation approaches, which is extremely dangerous for the cyber security systems as they always require a high level of safety. For instance, many researchers [263] [264] have proved the fact that it is possible to fool some of the most popular XAI explanation methods such as LIME and SHAP, which are also frequently deployed in the XAI application of cyber security areas. It is demonstrated that the explanations generating processes of those state-of-art XAI methods might be counter-intuitive. Other than that, in the practical industrial cyber security domains, such as XAI-enabled face authentication systems. Although in Section V, we have discussed several defensive methods against cyber threats targeting XAI systems, most defensive approaches focus on the protection of the performance of the prediction results of XAI models rather than the explanation results. However, for XAI-based cyber security systems, the explainability of the models is significant to maintain the transparency and efficiency of the entire system and prevent the cyber attacks as well.

#### 4) PRIVACY AND ETHICAL ISSUES

In addition to the aforementioned technical challenges, privacy and ethical issues are also crucial challenges when implementing XAI in cyber security. During the system life cycle, XAI models must explicitly take privacy concerns into account. It is commonly agreed that respecting every person's right to privacy is essential, especially in some very sensitive areas of cyber security, for instance, authentication, e-mails, and password. Moreover, XAI systems naturally fall within the general ethical concern of potential discrimination (such as racism, sexism, and ageism) by AI systems. In theory, identical biases may be produced by any AI model that is built using previously collected data from humans. It is important to take precautions to ensure that there is no discrimination, bias, or unfairness in the judgments made by the XAI system and the explanations that go along with them. The ethical bias of XAI systems should be eliminated in terms of justification as well as explainability, in particular in specific domains of cyber security applications. For privacy issues, because the data are gathered from security-related sources, the privacy and security-related concerns increase. Therefore, it is essential to guarantee that data and models are protected from adversarial assaults and being tampered with by unauthorized individuals, which means that only authorized individuals should be permitted access to XAI models.

### B. KEY INSIGHTS LEARNED FROM USING XAI FOR CYBER SECURITY

In this section, some key insights learned from using XAI for cyber security will be discussed based on the review in the above sections. The main insights for the XAI implementation in cyber security systems can be itemized as follows:

1) User trust and reliance should be satisfied. By offering explanations, an XAI system can increase end users' trust in the XAI-based cyber security system. Users of an XAI system can test their perception of the system's correctness and reliability. Users become dependent on the system as a result of their trust in the XAI-based cyber security system.
2) Model visualization and inspection should be considered. Cyber security experts could benefit from XAI system visualization and explainability to inspect model uncertainty and trustworthiness. Additionally, identifying and analyzing XAI model and system failure cases is another crucial component of model visualization and inspection.
3) Model tuning and selection are crucial factors to ensure the efficiency of the XAI model implemented in cyber security. Selecting different explanation approaches for distinct ML or DL algorithms in different cyber security tasks would influence the performance and explainability of XAI models significantly. Other than that, the tuning process of parameters and model structures of the established XAI model is another crucial consideration as well.
4) The model defense could be highlighted in particular for cyber security tasks as they are the main targets for cyber attackers. Especially for XAI-based cyber security mechanisms, the decision model, security data as well as the explanation process should be protected to prevent cyber threats.
5) Privacy awareness is another insight that XAI methods could provide for the cyber security system. Giving end users of cyber security systems a way to evaluate their data privacy is a significant objective in the application of XAI. End-users could learn through XAI explanations about what user data is used in algorithmic decision-making.

### C. FUTURE RESEARCH DIRECTIONS

#### 1) HIGH-QUALITY DATASETS

The quantity and quality of the available datasets have a significant impact on how well XAI methods work for the cyber security system, and the biases and constraints of the datasets used to train the models have an impact on how accurate the decisions and explanations are. On the other hand, as we discussed in the above sections, the existing available cyber security datasets could not reflect the most recent cyber attacks due to privacy and ethical issues. Data from real networks or the Internet typically contain sensitive information, such as personal or business details, and if made publicly available, they may disclose security flaws in the network from which they originated. Additionally, the imbalance of both volumes and features of the datasets would influence the establishment of the XAI-based cyber security system negatively as well. Therefore, the construction of both high-quality and up-to-date datasets available for XAI applications for cyber security could be a possible future research direction.

#### 2) TRADE-OFF BETWEEN PERFORMANCE AND EXPLAINABILITY

It is essential for cyber security experts to maintain the trade-off between performance and explainability aspects of the newly introduced XAI-enabled cyber security systems. It is noticeable that although for some self-explainable XAI approaches, for instance, Decision Tree, the model is quite transparent and users could understand the decision-making process easier, the performance of those approaches could not always be satisfying. On the other hand, the AI algorithms that now often perform best (for example, DL) are the least explainable, causing a demand for explainable models that can achieve high performance. Some researchers have exploited this area, including authors of [269] significantly reduce the trade-off between efficiency and performance by introducing XAI for DNN into existing quantization techniques. And authors of [270] demonstrated that the wavelet modifications provided could lead to significantly smaller, simplified, more computationally efficient, and more naturally interpretable models, while simultaneously keeping performance. However, there is a lack of research focusing on the trade-off of performance and explainability of XAI approaches applied in cyber security.

#### 3) USER-CENTERED XAI

The human understandability of XAI approaches has become the focus of some recent studies to find new potential for its application in areas of cyber security. As we mentioned in the above sections, user satisfaction with the generated explanation is a significant component of the XAI approaches to explainability evaluation. However, in areas of cyber security, the questionnaire and feedback of users are limited to some degree due to security concerns. Therefore, how to generate user-centered XAI systems for cyber security end users in terms of user understanding, user satisfaction, and user performance without violating the security issues could be a future research direction.

#### 4) MULTIMODAL XAI

Multimodal information of text, video, audio, and images in the same context can all be easily understood by people. The benefit of multimodality is its capacity to gather and combine important and comprehensive data from a range of sources, enabling a far richer depiction of the issue at hand. In some cyber security industrial areas, such as healthcare, medical decisions are primarily driven by a variety of influencing variables originating from a plurality of underlying signals and information bases, which highlights the need for multimodality at every stage. On the other hand, due to the application of XAI in these areas, multimodal XAI could be developed in near future.

#### 5) ADVERSARIAL ATTACKS AND DEFENSES

As we discussed in this review, although XAI could be applied in cyber security to prevent cyber attacks, the XAI model performance and explainability could be attacked as well. Other than that, the adversarial inputs to the sample data should be paid attention to as well. Some researchers [263] have already developed powerful tools to fool the state-of-art XAI methods including LIME and SHAP. However, although the cyber threats and corresponding defensive mechanisms focusing on the performance of AI models have been studied recently, the adversarial attacks and defenses against the explainability of XAI models still require further research.

#### 6) PROTECTION OF DATA

In cyber security areas, confidentiality and protection of data are significant issues as privacy and ethical issues are highlighted recently. For XAI-based systems, the situation is even more severe as both the decisions and the explanations related to users should be preserved. As a result, there is a conflict between using big data for security and safeguarding it. Data must be guaranteed to be safe from adversarial assaults and manipulation by unauthorized users and legitimate users should also be able to access the data. Therefore, the protection of data and generated explanations of XAI systems could be a future research direction as well.

## VII. CONCLUSION

XAI is a powerful framework to introduce explainability and transparency to the decisions of conventional AI models including DL and ML. On the other hand, cyber security is an area where transparency and explainability are required to defend against cyber security threats and analyze generated security decisions. Therefore, in this paper, we presented a comprehensive survey of state-of-art research regarding XAI for cyber security applications. We concluded the basic principles and taxonomies of state-of-art XAI models with essential tools, such as a general framework and available datasets. We also investigated the most advanced XAI-based cyber security systems from different perspectives of application scenarios, including XAI applications in defending against different categories of cyber attacks, XAI for cyber security in distinct industrial applications, and cyber threats targeting XAI models and corresponding defensive approaches. Some common cyber attacks including malware, spam, fraud, DoS, DGAs, phishing, network intrusion, and botnet were introduced. The corresponding defensive mechanisms utilizing XAI against them were presented. The implementation of XAI in various industrial areas namely in smart healthcare, smart financial systems, smart agriculture, smart cities, smart transportation, and Human-Computer Interaction were described exhaustively. Distinct approaches of cyber attacks targeting XAI models and the related defensive methods were introduced as well. In continuation to these, we pointed out and discussed some challenges, key insights and research directions of XAI applications in cyber security. We hope that this paper could serve as a reference for researchers, developers, and security professionals who are interested in using XAI models to solve challenging issues in cyber security domains.

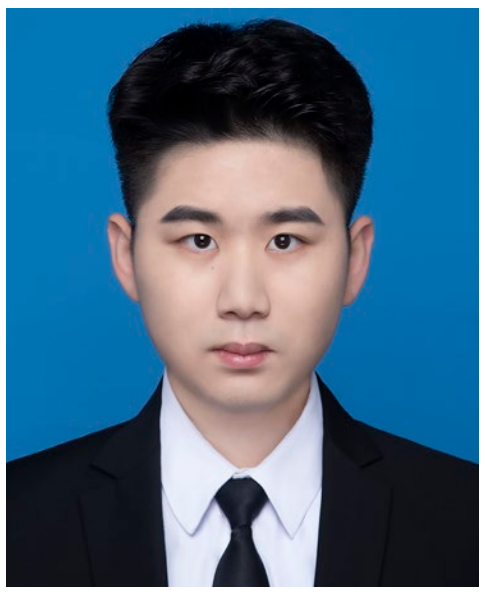

**ZHIBO ZHANG** received the Bachelor of Science degree in mechatronics engineering from Northwestern Polytechnical University, China, in 2021. He is currently pursuing a master's degree in electrical and computer engineering at Khalifa University, United Arab Emirates. His research interests focus on computer vision, cyber security, Explainable Artificial Intelligence, and Trustworthy Artificial Intelligence.

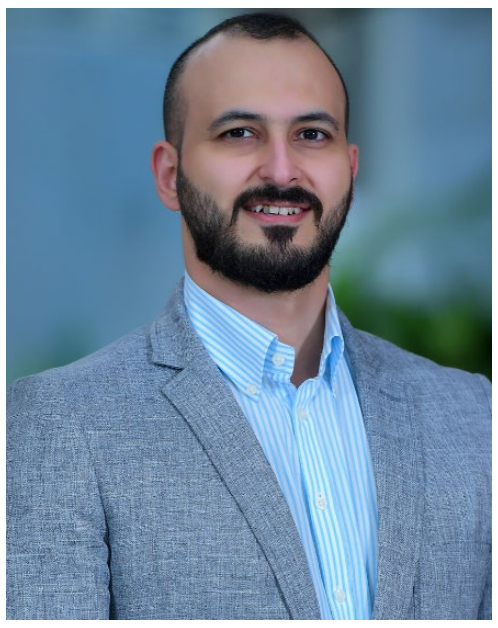

**HUSSAM AL HAMADI** (Senior Member, IEEE) studied computer engineering at Ajman University where he graduated in 2005. He spent the period between 2005 and 2010 working as a computer consultant and tutor in several governmental and private institutions to eventually joined the Khalifa University as a teaching assistant in 2010. He holds several international certificates in networking, business, and tutoring, like MCSA, MCSE, CCNA, CBP, and CTP. In 2017, he received his Ph.D. degree in computer engineering from Khalifa University, where he is currently a research scientist in their Center for Cyber-Physical Systems (C2PS). His research interests focus on applied security protocols for several systems like; software agents, SCADA, e-health systems and autonomous vehicles.

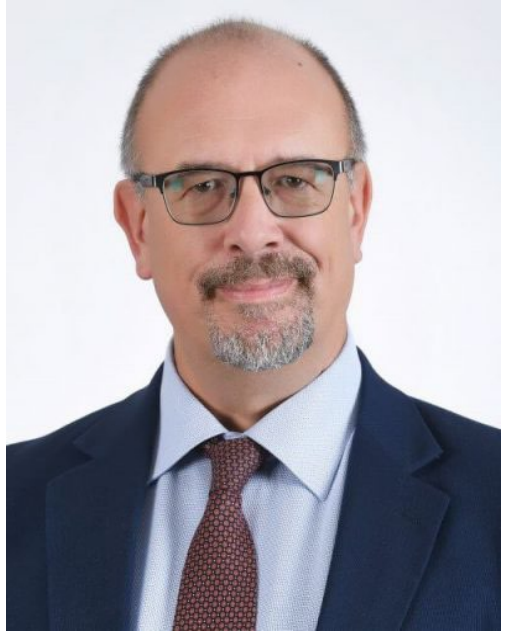

**ERNESTO DAMIANI** (Senior Member, IEEE) is currently a Full Professor with the Universitàdegli Studi di Milano, Italy, the Senior Director of the Robotics and Intelligent Systems Institute, and the Director of the Center for Cyber Physical Systems (C2PS), Khalifa University, United Arab Emirates. He is also the Leader of the Big Data Area, Etisalat British Telecom Innovation Center (EBTIC) and the President of the Consortium of Italian Computer Science Universities (CINI). He is also part of the ENISA Ad-Hoc Working Group on Artificial Intelligence Cybersecurity. He has pioneered model-driven data analytics. He has authored more than 650 Scopus-indexed publications and several patents. His research interests include cyber-physical systems, big data analytics, edge/cloud security and performance, artificial intelligence, and Machine Learning. He was a recipient of the Research and Innovation Award from the IEEE Technical Committee on Homeland Security, the Stephen Yau Award from the Service Society, the Outstanding Contributions Award from IFIP TC2, the Chester-Sall Award from IEEE IES, the IEEE TCHS Research and Innovation Award, and a Doctorate Honoris Causa from INSA-Lyon, France, for his contribution to big data teaching and research.

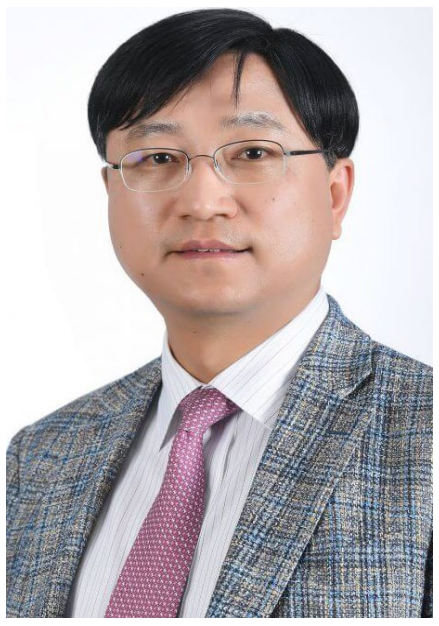

**CHAN YEOB YEUN** (Senior Member, IEEE) received the M.Sc. and Ph.D. degrees in information security from the Royal Holloway, University of London, in 1996 and 2000, respectively. After his Ph.D. degree, he joined Toshiba TRL, Bristol, U.K., and later became the Vice President at the Mobile Handset Research and Development Center, LG Electronics, Seoul, South Korea, in 2005. He was responsible for developing mobile TV technologies and related security. He left LG Electronics, in 2007, and joined ICU (merged with KAIST), South Korea, until August 2008, and then the Khalifa University of Science and Technology, in September 2008. He is currently a Researcher in cybersecurity, including the IoT/USN security, cyber-physical system security, cloud/fog security, and cryptographic techniques, an Associate Professor with the Department of Electrical Engineering and Computer Science, and the Cybersecurity Leader of the Center for Cyber-Physical Systems (C2PS). He also enjoys lecturing for M.Sc. cyber security and Ph.D. engineering courses at Khalifa University. He has published more than 140 journal articles and conference papers, nine book chapters, and ten international patent applications. He also works on the editorial board of multiple international journals and on the steering committee of international conferences.

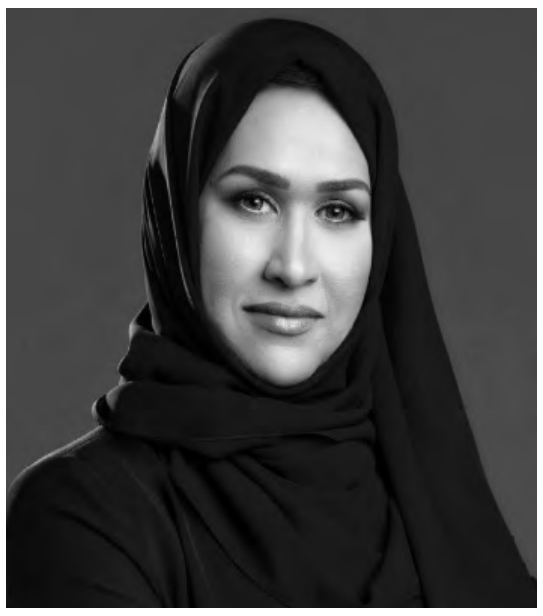

**FATMA TAHER** (Senior Member, IEEE) received the Ph.D. degree from the Khalifa University of Science, Technology and Research, United Arab Emirates, in 2014. She is currently the Assistant Dean of the College of Technological Innovation, Zayed University, Dubai, United Arab Emirates. She has published more than 40 articles in international journals and conferences. Her research interests are in the areas of signal and image processing, pattern recognition, Deep Learning, Machine Learning, artificial intelligence, medical image analysis, especially in detecting of the cancerous cells, kidney transplant, and autism. In addition to that, her researches are watermarking, remote sensing, and satellite images. She served as a member of the steering, organizing, and technical program committees of many international conferences. She has received many distinguished awards, such as the Best Paper Award of the first prize in the Ph.D. Forum of the 20th IEEE International Conference on Electronics, Circuits, and Systems (ICECS), the Ph.D. Forum, December 2013. And recently, she received the UAE Pioneers Award as the first UAE to create a computer-aided diagnosis system for early lung cancer detection based on the sputum color image analysis, awarded by H. H. Sheik Mohammed Bin Rashed Al Maktoum, November 2015. In addition to that, she received the Innovation Award at the 2016 Emirati Women Awards by H. H. Sheik Ahmed Bin Saeed Al Maktoum. She was the Chairman of Civil Aviation Authority and a Patron of Dubai Quality Group and L'Oréal-UNESCO for Women in Science Middle East Fellowship 2017. She is the Vice Chair of the IEEE UAE section and the Chair of the Education Committee in British Society, United Arab Emirates. She has served on many editorial and reviewing boards of international journals and conferences.